\documentclass[a4paper,11pt]{article}
\pdfoutput=1 
\RequirePackage{snapshot}  

\usepackage{jheppub} 

\usepackage{bm,amsmath,amssymb,slashed,graphicx,%
            enumerate,alltt,xspace,multirow,xcolor,mathrsfs}
\usepackage{fancyvrb}
\usepackage{booktabs}
\usepackage{graphicx}
\usepackage{subcaption}
\usepackage{xspace}
\usepackage[utf8]{inputenc}

\makeatletter
\g@addto@macro\bfseries{\boldmath}
\makeatother

\definecolor{labelkey}{rgb}{0,0.5,0.0}

\usepackage{listings}
\definecolor{darkgreen}{rgb}{0,0.4,0}
\definecolor{grey}{rgb}{0.5,0.5,0.5}

\newcommand{\CMW}{\text{CMW}}
\newcommand{\hint}[1]{\text{#1}}

\newcommand{\ariadne}{\texttt{Ariadne}\xspace}
\newcommand{\pythia}{\texttt{Pythia}\xspace}
\newcommand{\dire}{\texttt{Dire}\xspace}
\newcommand{\vincia}{\texttt{Vincia}\xspace}
\newcommand{\deductor}{\texttt{Deductor}\xspace}

\newcommand{\GeV}{\;\mathrm{GeV}}

\newcommand{\order}[1]{{\cal O}\left(#1\right)}
\newcommand{\as}{\alpha_s}
\newcommand{\MSbar}{\ensuremath{\overline{\text{MS}}}}

\newcommand{\state}{{\cal S}}
\newcommand{\cP}{{\cal P}}
\newcommand{\ptilde}{{\widetilde p}}
\newcommand{\etatilde}{{\widetilde \eta}}
\newcommand{\ztilde}{{\widetilde z}}
\newcommand{\evol}{\text{evol}}
\newcommand{\abar}{{\bar \alpha}}
\newcommand{\nc}{N_\text{\textsc{c}}}

\title{Logarithmic accuracy of parton showers: a fixed-order study}
\preprint{CERN-TH/2018-113}

\newcommand{\OXaff}{Rudolf Peierls Centre for Theoretical Physics, 1 Keble Road, Oxford OX1 3NP, UK}

\author[a]{Mrinal Dasgupta,}%
\author[b]{Fr\'ed\'eric A. Dreyer,}%
\author[c,d]{Keith Hamilton,}%
\author[d]{Pier Francesco Monni,}%
\author[d,*]{Gavin P.~Salam,\note[*]{On leave from CNRS, UMR 7589,
    LPTHE, F-75005, Paris, France and from \OXaff.}}%

\emailAdd{Mrinal.Dasgupta@manchester.ac.uk}
\emailAdd{fdreyer@mit.edu}
\emailAdd{keith.hamilton@ucl.ac.uk}
\emailAdd{pier.monni@cern.ch}
\emailAdd{gavin.salam@cern.ch}

\affiliation[a]{Consortium for Fundamental Physics, School of Physics and Astronomy, University of Manchester, Manchester M13 9PL, United
  Kingdom}
\affiliation[b]{Center for Theoretical Physics, Massachusetts
  Institute of Technology, Cambridge, MA 02139, USA}
\affiliation[c]{Department of Physics and Astronomy, University College London, London, WC1E 6BT, UK}
\affiliation[d]{CERN, Theoretical Physics Department, CH-1211 Geneva 23,
Switzerland}

\date{Received: date / Accepted: \today}

\abstract{
  We formulate some first fundamental elements of an approach for
  assessing the logarithmic accuracy of parton-shower algorithms based
  on two broad criteria: their ability to reproduce the singularity
  structure of multi-parton matrix elements, and their ability to
  reproduce logarithmic resummation results.
  We illustrate our approach by considering properties of two
  transverse-momentum ordered final-state showers, examining features
  up to second order in the strong coupling.
  In particular we identify regions where they fail to reproduce the
  known singular limits of matrix elements.
  The characteristics of the shower that are responsible for this also
  affect the logarithmic resummation accuracies of the shower, both in
  terms of leading (double) logarithms at subleading $N_C$ and
  next-to-leading (single) logarithms at leading $N_C$.
}


\begin{document}

\maketitle


\section{Introduction}
\label{sec:intro}

One of the most central and flexible tools in collider physics is the
general purpose Monte Carlo (GPMC) event
generator~\cite{Buckley:2011ms}, which simulates fully realistic
collider events.
Event generators involve various components to simulate physics at a
range of different scales:
for example ``hard'' scattering matrix elements describe the physics
occurring at the electroweak and TeV scales, the physics that
colliders are mainly built to probe.
Hadronisation models simulate the GeV-scale (``soft'') processes by
which quarks and gluons transform from and to the hadrons that are
collided and detected.
Parton shower algorithms, the subject of this article, provide the
crucial connection between the hard and soft scales, simulating mostly
strong-interaction physics across the two, three or even more orders
of magnitude of momenta that separate them.

GPMCs are extremely successful programs, able to reproduce much of the
data from CERN's Large Hadron Collider and its predecessors, sometimes to
quite high accuracy.
Part of this success is thanks to substantial progress in the past 20
years in improving the description of the hard scale, for example with
simultaneous matching to multiple tree-level matrix
elements~\cite{Mangano:2001xp,Catani:2001cc,Lonnblad:2001iq};
inclusion of next-to-leading (NLO) order corrections through the
MC@NLO~\cite{Frixione:2002ik,Alwall:2014hca} 
and POWHEG~\cite{Nason:2004rx,Frixione:2007vw,Alioli:2010xd} and other
\cite{Jadach:2015mza} methods;
and more recently merging of NLO corrections across processes of
varying jet multiplicity
simultaneously~\cite{Hoeche:2012yf,Hamilton:2012np,Frederix:2012ps},
and the incorporation of NNLO corrections in colour-singlet production
simulations~\cite{Hamilton:2013fea,Hoeche:2014aia,Alioli:2013hqa}.
Another important area of development contributing to the success of
today's GPMCs has been in their non-perturbative models, for example
for multiple interactions~\cite{Sjostrand:1987su,Butterworth:1996zw,Borozan:2002fk,Sjostrand:2004pf,Sjostrand:2004ef,Bahr:2008dy,Corke:2009tk,Corke:2011yy,Seymour:2013qka}, as well as their tuning to
data~\cite{Buckley:2009bj,Skands:2010ak}.

As GPMCs evolve towards higher accuracies, a number of factors make it
increasingly important to re-examine their parton shower component.
For example:
(1) the ability to match NNLO and higher-order calculations with
parton showers is to some extent limited by the fact that parton
showers do not reproduce the known structure of singularities that is
present in a NNLO calculation.
This is an issue especially for extensions of the MC@NLO method.
(2) The parton shower provides the initial conditions for
hadronisation, and the quality of a tune of hadronisation models can
well be affected by the quality of that initial condition.
This is especially true if one tunes the model predominantly using
data at one energy scale (e.g.\ LEP energies) and wishes to use the
same tune for much higher-energy processes (e.g.\ multi-TeV processes
at the LHC), where the parton shower is effectively providing much of
the extrapolation.
(3) There is an increasing trend towards exploiting information over
the full range of momenta between the hard scale and the hadronisation
scale, notably in jet substructure studies, using both hand-crafted
observables and machine learning~\cite{Larkoski:2017jix}.
Parton showers are the only tool with the flexibility to predict the
relevant dynamics for arbitrarily complex observables across that
range of scales.
(4) A number of experimental measurements are becoming limited by
systematic errors that have their origins in differences between
GPMCs, and one cannot help but wonder whether a better understanding
of parton showers could resolve this situation.
A notable example is the jet-energy-scale systematic uncertainty, for
which differences in quark and gluon fragmentation between different
GPMCs contribute significantly~\cite{Aad:2014bia,Khachatryan:2016kdb}.

There are several ongoing efforts to improve QCD parton showers, which can
be classified into three broad kinds of activity:
(1) developing novel showers that remain within the standard $1 \to 2$
or $2\to 3$ branching paradigms, for example the
\dire~\cite{Hoche:2015sya},
\vincia~\cite{Giele:2007di,Fischer:2016vfv} or \deductor~\cite{Nagy:2014mqa}
showers. 
(2) Incorporating more information about interferences, particularly
relevant for spin and colour degrees of freedom, e.g.\
Refs.~\cite{Nagy:2007ty,Nagy:2012bt,Nagy:2014mqa,Nagy:2015hwa,Martinez:2018ffw}.
(3) Incorporating higher-order splitting
functions~\cite{Jadach:2010aa,Jadach:2015mza,Jadach:2016zgk,Hoche:2017hno,Hoche:2017iem,Dulat:2018vuy,Li:2016yez}.
These efforts have brought significant innovations, however there
remains a need for a broader, systematic framework in which to think
about the question of parton-shower accuracy, so as to help frame and
guide such developments.
In particular to evaluate the advances brought by any single
improvement to a shower (e.g.\ higher-order splitting functions), one
needs to understand its interplay with the shower as a whole.
It is therefore the purpose of this article to sketch such a framework
and draw some first conclusions within it.

An important consideration in discussing the accuracy of parton
showers is that they may be used to calculate essentially any
observable.
This is to be contrasted with the situation for fixed order
calculations, where one selects a given observable, e.g.\ $n$-jet
kinematic distributions, and immediately knows which inputs are needed
for a given perturbative accuracy.
Here we suggest that a framework for discussing parton shower accuracy
should involve at least two core elements.

The first element of our proposed framework reflects the fact that
parton showers effectively generate an approximation to the tree-level
$n$-parton matrix elements for any $n$.
One can ask in what sense that approximation is correct.
Typically one would expect the parton shower to reproduce the matrix
element in a variety of ordered limits, where one or other kinematic
property of emissions is ordered from one emission to the next, e.g.\
$p_{\perp,n} \ll p_{\perp,n-1} \ll \ldots \ll p_{\perp,1} \ll Q$ (here
$p_{\perp,i}$ is the transverse momentum of particle $i$ and $Q$ is
the centre-of-mass energy).
Much is known about the properties of matrix elements in such ordered
limits, e.g.\
Refs.~\cite{Bassetto:1984ik,Dokshitzer:1992ip,Campbell:1997hg,Catani:1999ss}. 

We note that early, pioneering work on transverse-momentum ordered
showers with dipole-local
recoil~\cite{Gustafson:1987rq,Lonnblad:1992tz},\footnote{The shorthand
  term ``dipole shower'' appears to have different meanings for
  different groups, and so we avoid its use.}
did consider comparisons to matrix elements, both double-emission
matrix elements~\cite{Andersson:1991he} and, from the point of view of the colour
structure~\cite{Gustafson:1992uh}, multi-parton matrix
elements~\cite{Friberg:1996xc,Eden:1998ig}.
The lessons and algorithms derived from Ref.~\cite{Andersson:1991he},
about recoil prescriptions, were included in the \ariadne dipole
shower~\cite{Lonnblad:1992tz}, while those from
Ref.~\cite{Friberg:1996xc} were incorporated in a modification used in
Ref.~\cite{Eden:1998ig}.
We further discuss these articles below.

The second element of our framework makes a connection with
resummation.
Resummation accounts for logarithmically enhanced terms $\as^n L^m$ to all
orders, where $\as$ is the strong coupling and $L$ is the logarithm of
the ratio of any two physical scales.
Resummations are classified according to the specific set of dominant
and subdominant terms that are under control.
It is natural to ask what resummation accuracy will be obtained within
a given shower, for each observable where resummed results exist.
While this appears to be an observable-specific question, resummations
exist for large classes of
observables~\cite{Banfi:2004yd,Banfi:2014sua} and so in effect one can
ask questions about parton-shower resummation accuracy across all
observables within those classes.
Note that early work on showers~\cite{Marchesini:1983bm} gave
particular emphasis to the question of the scaling of particle
multiplicities with centre-of-mass energy, which is calculable
analytically.

There are natural connections between the two elements of our
framework.
For example, a failure to reproduce appropriate matrix-element limits
often occurs together with a failure to obtain a related resummation
accuracy for certain observables.
Nevertheless the two elements are also complementary: on one hand the
matrix-element conditions are relevant for observables of arbitrary
complexity, for which no resummation exists; on the other hand the
resummation conditions more immediately constrain aspects associated
with non-trivial virtual corrections, for example the scale of the
strong coupling.
The two elements are not necessarily exhaustive in terms of the types
of requirements one may ask of showers, but as we will see they are
already informative and constraining.

To make our discussion concrete, it will be helpful to examine the
case of specific showers.
We will consider two transverse-momentum ordered showers with
dipole-local recoil: the {\tt Pythia}
 shower~\cite{Sjostrand:2004ef}, and the {\tt Dire}
shower~\cite{Hoche:2015sya}, restricting our attention to massless
final-state splittings.
The algorithms underlying these showers are described in
section~\ref{sec:shower-summary}.
Then in section~\ref{sec:singularity-structure} we will outline how
to usefully classify the ability of showers to reproduce $n$-parton
matrix elements.
This will reveal issues already at the two-emission level. 
Section~\ref{sec:fixed-order} will then show how these issues connect
with the question of logarithmic resummation accuracy.
Again a second-order analysis will be sufficient to highlight the main
features.

\section{Parton showering and our choice of algorithms}
\label{sec:shower-summary}

There are many parton showers being used and under further development
today.
They generate emissions in a sequence according to a kinematic
ordering variable.
One way of classifying showers is based on their specific
choice of ordering variable.
The most common choice is to order emissions in transverse momentum
and all three major Monte Carlo programs have such a shower:
{\tt Pythia}~\cite{Sjostrand:2006za,Sjostrand:2014zea} uses the shower of
Ref.~\cite{Sjostrand:2004ef}, {\tt Sherpa}~\cite{Gleisberg:2008ta} uses the
shower of Ref.~\cite{Schumann:2007mg} and {\tt Herwig} has as an option the
shower of Ref.~\cite{Platzer:2009jq}.
Other transverse-momentum showers include
{\tt Vincia}~\cite{Giele:2007di,Fischer:2016vfv}, available in {\tt Pythia}, and
{\tt Dire}~\cite{Hoche:2015sya}, available for both {\tt Pythia} and {\tt Sherpa}.
Another class of shower orders emissions in
angle~\cite{Marchesini:1983bm,Gieseke:2003rz,Reichelt:2017hts} and is
used mainly in the {\tt Herwig} Monte Carlo programs~\cite{Bellm:2015jjp}.
Finally there is the {\tt Deductor} shower work~(Ref.~\cite{Nagy:2014mqa} and
references therein), which considers an ordering variable that is
related to emission time, but also discusses transverse-momentum
ordering. Its code is standalone.

A further important distinguishing feature of each parton shower is
the way in which the recoil associated with the emitted parton at a
given evolution step is absorbed by other particles in the event. Some
shower algorithms rely on a \emph{local} scheme, in which the recoil is
shared among the two colour-connected partners of the
emission. Another approach is to use a \emph{global} scheme, which
distributes the recoil among all other particles in the event.

It is beyond the scope of this article to consider all of these showers.
We rather choose to concentrate on two of them: (1) the {\tt Pythia} shower
on the grounds that it is today's most extensively used shower;
and (2) the {\tt Dire} shower~\cite{Hoche:2015sya}, on the grounds that it
is the only shower explicitly available in two Monte Carlo simulation
programs ({\tt Pythia}~\cite{Sjostrand:2014zea} and
{\tt Sherpa}~\cite{Gleisberg:2008ta}) and that it is being used as a basis
for the inclusion of higher-order splitting
kernels~\cite{Hoche:2017hno,Hoche:2017iem,Dulat:2018vuy}.
Both are transverse-momentum ordered and use recoil that is kept local
within colour dipoles.\footnote{Neither of these showers claims NLL
  accuracy. For example the Pythia manual states ``While the final
  product is still not certified fully to comply with a NLO/NLL
  standard, it is well above the level of an unsophisticated LO/LL
  analytic calculation.''~\cite{Sjostrand:2006za}.}

To help our discussion it is useful to give a summary of the
ingredients of common parton shower algorithms.
We use $\state_n$ to denote a specific kinematic state with $n$
partons.
The probability $P(\state_n,v)$ of finding that state is a function of
a kinematic ordering variable $v$.
A first key component of a parton shower algorithm is a differential
equation for the evolution of that probability as the ordering
variable is decreased:
\begin{equation}
  \label{eq:baseEvolutionEquation}
  \frac{dP(\state_n, v)}{d\ln 1/v} = - f(\state_n, v) P(\state_n, v)\,.
\end{equation}
The second component of the parton shower algorithm is a kinematic
mapping from the state $\state_n$ to an $n+1$-particle state
$\state_{n+1}$.
The map is a function of the ordering variable, 
the choice of the partons involved in the branching and two
additional kinematic variables which we call $z$ and $\phi$.
Insofar as we deal with showers with recoil that is local to the
colour dipole that is splitting, one should choose two partons, which
we label $i$ and $j$.
We write the mapping as
\begin{equation}
  \label{eq:baseSplittingEquation}
  \state_{n+1} = {\cal M}(\state_n, v; i,j, z,\phi)\,.
\end{equation}
This kinematic map has an associated ``splitting'' weight function
$d\cP ({\state_n}, v; i,j,z,\phi)$, which governs the relative
probabilities of the different possible new states and which can be
conveniently normalised so as to relate it to the $f({\state_n},v)$
function of Eq.~(\ref{eq:baseEvolutionEquation}),
\begin{equation}
  \label{eq:mappingProb}
  f(\state_n, v) = \sum_{i,j} \int dv' dz d\phi \,
  \frac{d\cP ({\state_n}, v'; i,j,z,\phi)}{dv' dz d\phi} \delta(\ln v'/v)\,.
\end{equation}
Eq.~(\ref{eq:baseEvolutionEquation}) encodes the virtual contributions
associated with maintaining the system in state $\state_n$.
Eq.~(\ref{eq:mappingProb}) states that virtual and real contributions
should be equal (aside the opposite sign), i.e.\ that probability is
conserved, which is also referred to as unitarity.
In suitable soft and/or collinear limits (we return to this in
section~\ref{sec:singularity-structure}) one expects the splitting
weight function to be closely related to the ratio of $n+1$ and
$n$-parton matrix elements and phase space.
Schematically, one might write this as
\begin{equation}
  \label{eq:2}
  \sum_{i,j} d\cP({\state_n}, v; i,j, z,\phi) 
  \simeq \frac{d\Phi_{n+1}}{d\Phi_{n}}
  \frac{|M^2(\state_{n+1})|}{|M^2(\state_{n})|}\,.
\end{equation}
For typical dipole showers, the sum over $i$ runs over all emitting
particles and $j$ over all colour connected partners, at most two for
each $i$ at leading colour. 
Colour factors and the relevant factor of the strong coupling are
included in $d\cP$.

The difference between one shower and another lies not just in the
choice of kinematic ordering variable $v$, but also in the mapping
function $\cal M$ and the splitting weight function $\cP$.
For a given colour dipole $ij$, the showers that we consider here
separate the phase space into a region that is predominantly collinear
to $i$ and another that is predominantly collinear to $j$.

Note that for the purpose of this article we will only consider
final-state showers, with massless partons.

\subsection{Pythia $p_t$-ordered shower}
\label{sec:pythia-summary}

{\tt Pythia}'s transverse-momentum ordered shower~\cite{Sjostrand:2004ef}, is
the default option of the {\tt Pythia8} program~\cite{Sjostrand:2014zea}
and was also available in {\tt Pythia6}~\cite{Sjostrand:2006za}.
The exposition that follows, restricted to its final-state branching
elements, is based on Ref.~\cite{Sjostrand:2004ef} and inspection of
the {\tt Pythia8} code, version 8.266.

The ordering variable $v$ is a transverse momentum, which is referred
to as $p_{\perp,\text{evol}}$,
\begin{equation}
  \label{eq:1}
  v \equiv p_{\perp,\text{evol}}\,.
\end{equation}
The map ${\cal M}(\state_n, v; i,j, z,\phi)$  takes massless
pre-branching momenta $\ptilde_i$ and $\ptilde_j$ and constructs post
branching momenta $p_i$, $p_j$ and $p_k$, corresponding to a
branching $\tilde p_i \to p_i + p_k$ with spectator particle $j$
taking longitudinal recoil to ensure momentum conservation.
It is useful to define intermediate variables
\begin{equation}
  \label{eq:py8-vars}
  \rho_{\perp,\text{evol}}^2
  = \frac{p_{\perp,\text{evol}}^2}{(\ptilde_i+\ptilde_j)^2}\,,
  \qquad
  y = \frac{\rho_{\perp,\text{evol}}^2}{z(1-z)}\,,
  \qquad
  \tilde z
  = \frac{\left(1-z\right)\left(z^{2}-\rho_{\perp{\text{evol}}}^{2}\right)}{z\left(1-z\right)-\rho_{\perp{\text{evol}}}^{2}}\,,
\end{equation}
and the Catani--Seymour~\cite{Catani:1996vz} style dipole map is then
defined by
\begin{subequations}
  \label{eq:Py8-map}
  \begin{align}
    p^\mu_{i}\,=\, & \ztilde\,\ptilde^\mu_{i}+y\left(1-\ztilde\right)\ptilde^\mu_{j}+k^\mu_{\perp}\,,\label{eq:Py8-Catani-Seymour-Moms-p_b}\\
    p^\mu_{k}\,=\, & \left(1-\ztilde\right)\ptilde^\mu_{i}+y \ztilde\ptilde^\mu_{j}-k^\mu_{\perp}\,,\label{eq:Py8-Catani-Seymour-Moms-p_c}\\
    p^\mu_{j}\,=\, & \left(1-y\right)\ptilde^\mu_{j}\,.\label{eq:Py8-Catani-Seymour-Moms-p_r}
  \end{align}
\end{subequations}
Here $k^\mu_{\perp}$ is defined as 
\begin{equation}
  \label{eq:Py8-kperp}
  k^\mu_{\perp} = \sqrt{\ztilde(1-\ztilde)\, y \,(\ptilde_i + \ptilde_j)^2} \left[
    \widehat{k}^\mu_{\perp,1}\cos\phi+\widehat{k}^\mu_{\perp,2}\sin\phi
  \right]\,,
\end{equation}
where $\widehat{k}^\mu_{\perp,1}$ and $\widehat{k}^\mu_{\perp,2}$ are
four-vectors that are orthogonal to each other as well as to $\ptilde^\mu_i$,
$\ptilde^\mu_j$, and that satisfy $\widehat{k}_{\perp,1}^2 =
\widehat{k}_{\perp,2}^2 = -1$.
Eqs.~(\ref{eq:Py8-map})--(\ref{eq:Py8-kperp}) imply
\begin{equation}
  \label{eq:Py8-kperp-zrho}
  |k_{\perp}^2| =
  \frac{\left(z^{2}-\rho_{\perp{\text{evol}}}^{2}\right)\left(\left(1-z\right)^{2}-\rho_{\perp{\text{evol}}}^{2}\right)}%
  {\left(z\left(1-z\right)-\rho_{\perp{\text{evol}}}^{2}\right)^{2}}\,
  p_{\perp,\text{evol}}^2 \,.
\end{equation}
The squared transverse momentum $|k_{\perp}^2|$ that is assigned to
the emission coincides with the ordering variable
$p_{\perp,\text{evol}}^2$ when
$z, (1-z) \gg \rho_{\perp,\text{evol}}$, i.e.\ in the collinear
limit.
The map only exists for
\begin{equation}
  \label{eq:Py8-zlims}
  \rho_{\perp,\text{evol}} \le z \le 1-\rho_{\perp,\text{evol}}\,,
\end{equation}
and, at the edges of this range, $|k_{\perp}|$ vanishes even for finite
$p_{\perp,\text{evol}}$.

If parton $i$ is a quark, only a $q \to qg$ branching is possible, and
the quark is colour connected to only one other particle in the event
(the spectator $j$). The splitting weight function in this case is
given by
\begin{equation}
d\cP_{q\rightarrow
  qg}\,=\,
  \frac{\as(p_{\perp,\text{evol}}^2)}{2\pi}\,
  \frac{dp_{\perp{\text{evol}}}^{2}}{p_{\perp{\text{evol}}}^{2}}\,dz\,\frac{d\phi}{2\pi}\,
  C_{F}\,\left(\frac{1+z^{2}}{1-z}\right)\,.
\label{eq:Py8-Ingredients-q-qg-branching-prob}
\end{equation}
Note the use of the evolution variable $p_{\perp,\text{evol}}^2$ in
the scale of $\as$, rather than the kinematic quantity
$|k_{\perp}^2|$.
If parton $i$ is a gluon, both $g \to gg$ and $g \to q\bar q$
branchings are possible and for each of them, the shower takes into
account two colour connections, assigning equal weights to each.
The $g\to gg$ splitting weight is
\begin{equation}
  d\mathcal{P}_{g\rightarrow gg}%
  \,=\,\frac{\as(p_{\perp,\text{evol}}^2)}{2\pi}\,
  \frac{dp_{\perp{\text{evol}}}^{2}}{p_{\perp{\text{evol}}}^{2}}\,dz\,\frac{d\phi}{2\pi}\,
  \frac{C_{A}}{2}\,\left[\frac{1+z^{3}}{1-z}\right]\,.
  \label{eq:Py8-Ingredients-g-gg-branching-prob}
\end{equation}
The usual $P_{gg}$ splitting function is reconstructed from this,
together with its $1/2!$ symmetry factor, when one considers that each
gluon splits separately in each of the two dipoles to which it
belongs\footnote{i.e.\ in a $q g \bar q$ system,
  Eq.~(\ref{eq:Py8-Ingredients-g-gg-branching-prob}) applies to
  the gluon splitting in the $qg$ dipole and the gluon splitting in
  the $g\bar q$ dipole, each of which carries a $C_A/2$ factor.} and that for a $\ptilde_i \to p_i + p_k$
splitting, observables do not distinguish between gluon $p_i$ and
gluon $p_k$, which provides an implicit symmetrisation of
$z \leftrightarrow 1-z$.
For the $g \to q\bar q$ case, the following is used
\begin{equation}
  d\mathcal{P}_{g\rightarrow q\bar{q}}\,
  =\,\frac{\as(p_{\perp,\text{evol}}^2)}{2\pi}\,
    \frac{dp_{\perp{\text{evol}}}^{2}}{p_{\perp{\text{evol}}}^{2}}\,dz\,\frac{d\phi}{2\pi}\,
     \frac{n_{f}T_{R}}{2} \, \mathcal{D}\, \left[1-2\ztilde\left(1-\ztilde\right)\right]\,,
    \label{eq:Py8-Ingredients-g-qq-branching-prob}
\end{equation}
where again the gluon splitting occurs with this weight separately in
each of the two dipoles to which it belongs.
The factor $\mathcal{D}$ is,
\begin{equation}
  \label{eq:5}
  \mathcal{D} = (1-x)^2(1+x)\,, \qquad x \equiv
  \frac{(p_i+p_k)^2}{(\ptilde_i + \ptilde_j)^2}\,.
\end{equation}
Such that in the collinear, limit $\mathcal{D} =1$ and $\ztilde = z$.
The {\tt Pythia} shower has the option of using the
CMW~\cite{Catani:1990rr} scheme for the coupling
\begin{equation}
  \label{eq:3}
  \as^{\CMW}(p_{\perp,\text{evol}}^2) =
  \as^{\MSbar}(p_{\perp,\text{evol}}^2) \left(1 +
    \frac{\as^{\MSbar}(p_{\perp,\text{evol}}^2)}{2\pi} 
    K\right)\,,\qquad K = \left(\frac{67}{18} -
    \frac{\pi^2}{6}\right)C_A - \frac{10}{9} T_R n_f\,,
\end{equation}
in the soft-enhanced parts of the splitting functions, 
which is one key element of NLL resummations.

\subsection{Dire shower}
\label{sec:dire-summary}

The {\tt Dire} transverse-momentum ordered shower~\cite{Hoche:2015sya}, is
available for both the {\tt Sherpa}~\cite{Gleisberg:2008ta} and the {\tt Pythia8}
generation frameworks.

The {\tt Dire} ordering variable is once again a (squared) transverse momentum
type variable and is called $t$,
\begin{equation}
  \label{eq:Dire-v}
  v \equiv \sqrt{t}\,,
\end{equation}
together with the splitting variables $z$ and $\phi$.
To construct the final-state kinematic map one defines
intermediate variables
\begin{equation}
  \label{eq:Dire-intermediate}
  \kappa^2 = \frac{t}{(\ptilde_i + \ptilde_j)^2}\,,\qquad
  y = \frac{\kappa^2}{1-z}\,,\qquad
  \ztilde = \frac{z-y}{1-y}\,.
\end{equation}
The {\tt Dire} map then has the identical form to the {\tt Pythia} map, i.e.\
using Eqs.~(\ref{eq:Py8-map}) and (\ref{eq:Py8-kperp}) but with the {\tt Dire}
expressions for $y$ and $\ztilde$.
The kinematic squared transverse momentum, expressed in terms of the
original splitting variables, is
\begin{equation}
  \label{eq:Dire-kt2}
  |k_{\perp}^2| = (1-z)\, \frac{z(1-z) - \kappa^2}{(1-z - \kappa^2)^2}\, t\,.
\end{equation}
In the soft-collinear limit, with $1-z \ll 1$ and $\kappa \ll 1-z$,
this reduces to $|k_{\perp}^2| = t$, i.e.\ the ordering variable is
identical to the squared emitted transverse momentum, as in the case
of the {\tt Pythia} shower.
The limits on $z$ for a given value of $t$ are dictated by the
requirement of positivity of the right-hand side of
Eq.~(\ref{eq:Dire-kt2}), and give
\begin{equation}
  \label{eq:Dire-z-limits}
  \frac12 - \sqrt{\frac14 - \kappa^2} \le z \le \frac12 + \sqrt{\frac14 - \kappa^2}\,.
\end{equation}
For small $\kappa$, this becomes $\kappa^2 \le z \le 1 - \kappa^2$.
Note that this scales differently from the {\tt Pythia} case,
Eq.~(\ref{eq:Py8-zlims}), the
consequences of which will be discussed below.
The {\tt Dire} splitting weight functions are
\begin{subequations}
  \begin{align}
    d\cP_{q\rightarrow
    qg}\,&=\,
           \frac{\as(t)}{2\pi} \,
           \frac{dt}{t}\,dz\,\frac{d\phi}{2\pi}\,
           C_{F}\,\left[
           2\frac{1-z}{(1-z)^2 + \kappa^2} - (1+z)
           \right]\,,
           \label{eq:Dire-Ingredients-q-qg-branching-prob}
    \\
    d\mathcal{P}_{g\rightarrow gg}%
    \,&=\,\frac{\as(t)}{2\pi}\,
        \frac{dt}{t}\,dz\,\frac{d\phi}{2\pi}\,
        \frac{C_{A}}{2}\,\left[
        2\frac{1-z}{(1-z)^2 + \kappa^2}
        - 2 + z(1-z)
        \right]\,,
        \label{eq:Dire-Ingredients-g-gg-branching-prob}
    \\
    d\mathcal{P}_{g\rightarrow q\bar{q}}\,
         &=\,\frac{\as(t)}{2\pi}\,
           \frac{dt}{t}\,dz\,\frac{d\phi}{2\pi}\,
           \,\frac{n_{f}T_{R}}{2}\,\left[1 - 2z(1-z)\right]\,.
           \label{eq:Dire-Ingredients-g-qq-branching-prob}
  \end{align}
\end{subequations}
As in the case of {\tt Pythia} the gluon splittings apply once for
each of the two dipoles to which the gluon belongs.\footnote{The $z$
  structure of the $d\mathcal{P}_{g\rightarrow gg}$ formula that we
  quote is given by the sum of the \texttt{fsr\_qcd\_G2GG1::calc} and
  \texttt{fsr\_qcd\_G2GG2::calc}, functions in the {\tt Dire} 2.001
  code. Analogously for the other splitting functions.}
The {\tt Dire} shower (like {\tt Pythia}) has the option of using the
CMW scheme in the coupling, i.e.\ $\as^\text{CMW}(t)$.
One key difference of {\tt Dire} relative to the {\tt Pythia}
case is the modification of the soft divergence,
\begin{equation}
  \label{eq:pythia-to-dire-splittings}
  \frac1{1-z} \to \frac{1-z}{(1-z)^2 + \kappa^2}\,.
\end{equation}
This introduces an effective cutoff of the soft divergence when
$1-z \sim \kappa$, to be contrasted with the actual limit in the
kinematic map of $1-z \gtrsim \kappa^2$.
Thus the {\tt Pythia} and {\tt Dire} showers both effectively cut off the
divergence for $1-z \sim \kappa$, but {\tt Pythia} implements this through
the kinematic map, while {\tt Dire} does so through the splitting
functions.
Physically this cutoff is situated around zero rapidity in a frame in
which the dipole is at rest: effectively only one of the dipole's two
partons radiates in an $i \to i+k$ splitting, and the radiation fills
the associated hemisphere in the dipole centre of mass.

\section{Singularity structure of resulting matrix elements}
\label{sec:singularity-structure}

For each emission $i$ in an $n$-parton matrix element there are two
kinds of singularity, a soft singularity when parton $i$'s energy goes
to zero and a collinear singularity when its angle with respect to any
other parton goes to zero.
One has considerable freedom in what two variables one uses to
describe these two singularities.
For example one may use pair-invariant mass and energy, angle and
energy, transverse momentum and angle, etc.

The very minimal expectation for a parton shower is that it reproduces
the matrix element for any single-emission configuration with one or
two singularities: i.e.\ in the collinear and soft limit, with two
singularities;
in the collinear and non-soft limit, with one singularity
corresponding to the DGLAP splitting
functions~\cite{Gribov:1972ri,Altarelli:1977zs,Dokshitzer:1977sg};
and in the soft non-collinear limit, reproducing the
Bassetto-Ciafaloni-Marchesini eikonal emission
formulas~\cite{Bassetto:1984ik}.

Once one considers more than one emission, one reasonable expectation
is to control the leading singularity of the squared amplitude for any
number of emissions, that is to correctly reproduce the divergence of
the matrix element in configurations where each emission triggers two
singularities relative to a parent configuration without that
emission.
For example, the leading singularity for such a configuration might
involve two emissions with disparate values both of their transverse
momenta and angles.\footnote{For the purposes here, angles are
  understood to be defined with respect to the more energetic particle
  to which they are closest in angle.}
This is closely connected with the reproduction of leading double
logarithms.

Among the elements needed to reproduce subleading-logarithmic
corrections, one might also require that when there are one or more
emissions that each trigger only one singularity (rather than two),
the matrix element is still correctly reproduced.
For example, multiple hard emissions that are ordered in angle should
reproduce the DGLAP anomalous dimensions (this was discussed and
eventually established for dipole type showers in
Refs.~\cite{Dokshitzer:2008ia,Nagy:2009re,Skands:2009tb}).
Such a limit also involves spin-correlations, for which an algorithm
has long been known~\cite{Collins:1987cp,Knowles:1988hu}, though only
some showers make use of it.

A potentially delicate configuration with regards to this condition
occurs when emissions have commensurate values of the ordering
variable, but disparate values of a complementary kinematic variable.
For example, in angular-ordered showers, it is known that the matrix
element is not correctly reproduced for multiple emissions with
commensurate angles but strongly ordered energies.
As a result non-global single-logarithmic terms are not correctly
reproduced~\cite{Banfi:2006gy}.

One can study the reproduction of singularities both exactly
and in a leading $N_\text{\textsc{c}}$ limit.
For example the case of commensurate angles and disparate energies is
especially challenging beyond leading $\nc$, even within dedicated resummation
approaches, and only one complete answer is known~\cite{Hatta:2013iba}. 

For what follows, we will work in a limit where all emissions are
soft relative to the centre-of-mass energy.
They may be strongly ordered both in transverse momentum and angle, or
only in the latter.
This limit will already prove to be illuminating.

\subsection{Single-emission case}
\label{sec:single-emission-case}

To reproduce the leading double logarithms, the requirement for
the single-emission pattern is that in the limit where an emission has a
small energy with respect to the parent system (soft limit) and it has
a small angle relative to another parton (collinear limit), the
corresponding matrix element times phase space should be reproduced,
i.e.\ 
\begin{equation}
  \label{eq:one-emission-correct}
  d\cP = \frac{2 C \as(p_\perp^2)}{\pi} \frac{dp_\perp}{p_\perp} d\eta
\end{equation}
where $C$ is the colour factor of the emitting parton, $\eta$ is a
rapidity with respect to the emitting parton and $p_\perp$ is a
transverse momentum with respect to it.
This is straightforwardly reproduced by our selected showers.

For single-logarithmic accuracy, it is essential to also reproduce the
emission pattern in the hard collinear region and that in the soft
large-angle region.
The former is straightforward to verify from the equations in
section~\ref{sec:shower-summary}, after taking into account the
symmetrisation over $z \leftrightarrow (1-z)$ and the fact that gluons
radiate separately as part of each of two dipoles.
It is therefore the soft large-angle region that we examine here.
To discuss the soft large-angle region it is helpful to use
(pseudo)rapidity, $\eta = -\ln \tan \theta/2$ and physical transverse
momentum, $|p_\perp|$ of the emission to parametrise its phase space.
A variety of definitions can be constructed for the physical
transverse momentum, but in the soft limit, for emission from a single
colour dipole, all sensible ones will coincide.\footnote{E.g.\ for a
  massless emission $p$ from a dipole between massless particles $P_1$
  and $P_2$, one can define
  $p_\perp^2 = 2 \frac{(P_1.p) (P_2.p)}{P_1.P_2}$.}

For concreteness, consider a $q\bar q$ dipole of mass $Q$, in its
centre-of-mass frame, with the quark $q$ along the $z$ axis.
First we consider branching of the quark.
Using Eq.~(\ref{eq:Py8-map}), the emitted gluon has
pseudorapidity:
\begin{equation}
  \label{eq:eta}
  \eta = \ln \frac{(1-\ztilde) Q}{|k_\perp|}\,,
\end{equation}
where $k_\perp$ in Eq.~(\ref{eq:Py8-map}) coincides with our
$p_\perp$.
The {\tt Pythia} mapping in the soft limit,
$1-z \ll 1$, gives
\begin{equation}
  \label{eq:eta-kt-py8}
  \hint{{\tt Pythia}:}\qquad
  \eta = \frac12 \ln \left[
    \frac{(1-z)^2}{\rho_{\perp,\text{evol}}^2}-1\right]\,,\qquad
  |p_\perp^2| = p_{\perp,\evol}^2 \left(1 - \frac{\rho_{\perp,\evol}^2}{(1-z)^2}\right)\,.
\end{equation}
For $z/\rho > \sqrt{2}$, $\eta$ is positive.
It is instructive to examine the contour in the $\eta, \ln |p_\perp|$
plane that is covered for a given value of the ordering variable,
$v \equiv p_{\perp,\text{evol}}$, together with the splitting function weight
differentially along that contour: 
\begin{align}
  \label{eq:py8-kt-eta-P}
  \hint{{\tt Pythia}:}\qquad
  |p_\perp^2| = p_{\perp,\evol}^2 \left(
    \frac{e^{2\eta}}{1 + e^{2\eta}}\right)\,,
  \qquad
  d\cP_{q\to qg} = \frac{2 \as(p_{\perp,\evol}^2) C_F}{\pi} \frac{d p_\perp}{p_\perp} d\eta 
  \left(\frac{e^{2\eta}}{1 + e^{2\eta}}\right)\,,
\end{align}
where we have dropped the $d\phi/2\pi$ factor for compactness.
For large positive values of $\eta$, $|p_\perp| = p_{\perp,\evol}$ and
the splitting weight is independent of $\eta$.
For large negative values of $\eta$,
$|p_\perp| = e^{-|\eta|} p_{\perp,\evol}$ and the splitting weight is
suppressed, but non-zero.
The splitting of the anti-quark $\bar q$ yields similar results, but
with $\eta \to - \eta$.
If one ignores the running of the coupling, the sum of the $q$ and
$\bar q$ splittings yields
\begin{equation}
  \label{eq:py8-P-sum}
  \hint{{\tt Pythia} (ignoring running):}\qquad 
  d\cP_{q\to qg} + d\cP_{\bar q\to \bar q g}
  = \frac{2 \as C_F}{\pi} \frac{d p_\perp}{p_\perp} d\eta\,.\end{equation}
This has a uniform distribution in rapidity, which is the correct result
for soft-gluon emission from a dipole.
The analogue of Eqs.~(\ref{eq:eta-kt-py8}), (\ref{eq:py8-kt-eta-P}) for {\tt Dire} is
\begin{equation}
  \label{eq:dire-kt-eta-P}
  \hint{{\tt Dire}:}\qquad
  \eta=\frac{1}{2}\ln\left[\frac{(1-z)^{2}}{\kappa^{2}}\right]\,,
  \qquad
  |p_{\perp}^{2}|=t\,,
  \qquad
  d\cP_{q\to qg}=\frac{2\as(t)C_{F}}{\pi}\frac{dp_{\perp}}{p_{\perp}}d\eta
  \left(\frac{e^{2\eta}}{1+e^{2\eta}}\right)\,.
\end{equation}
{\tt Dire} and {\tt Pythia} therefore have identical rapidity distributions for
soft emission from one side of a dipole, however in {\tt Dire} the emission
always has $|p_\perp^2| = t$, unlike the {\tt Pythia} case.
This means that the sum of quark and anti-quark splittings has a
simple weight even taking into account running coupling effects:
\begin{equation}
  \label{eq:dire-P-sum}
  \hint{{\tt Dire}:}\qquad 
  d\cP_{q\to qg} + d\cP_{\bar q\to \bar q g}
  = \frac{2 \as(|p_\perp^2|) C_F}{\pi} \frac{d p_\perp}{p_\perp} d\eta\,.
\end{equation}

\begin{figure}
  \centering
  \begin{subfigure}[t]{0.51\textwidth}
    \includegraphics[width=\textwidth,height=0.26223\textheight]{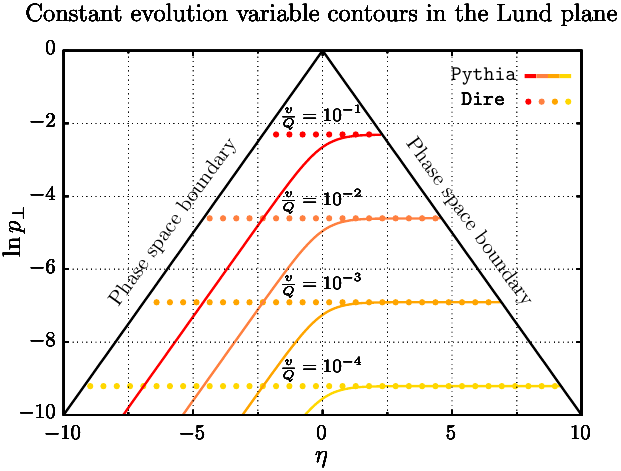}%
    \caption{}
    \label{fig:ME-1-parton-const-v-contours}
  \end{subfigure}\hfill%
  \hfill
  \begin{subfigure}[t]{0.49\textwidth}
    \includegraphics[width=\textwidth]{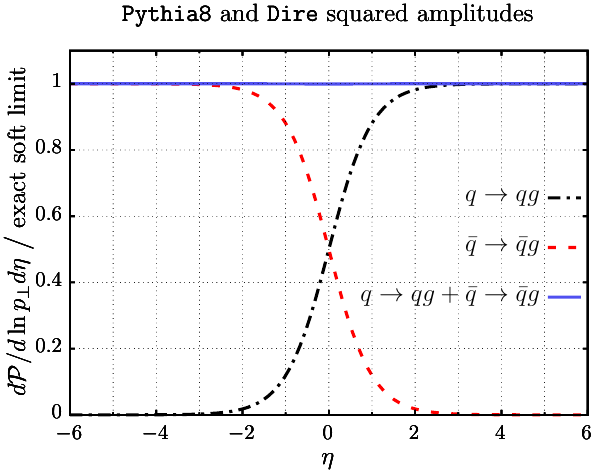}
    \caption{}
    \label{fig:ME-1-parton-dipole-partition-weights}
  \end{subfigure}
  \caption{(a) The accessible contour of emissions in the
    $\eta{-}\ln p_\perp$ (``Lund'' \cite{Andersson:1988gp}) plane
    for fixed values of the 
    ordering variable $v$, for splittings of a
    right-going quark, shown for both the \pythia and
    \dire shower kernels.
    The phase-space boundary is sometimes alternatively described as
    the hard-collinear limit.
    (b) The splitting weights associated with emissions as a
    function of rapidity from the (right-going) quark and (left-going)
    anti-quark, normalised so as to be $1$ in the soft-collinear region.
    This holds for both \pythia and \dire.
    The weights are independent of $v$, as long as $v/Q \ll
    e^{-|\eta|}$, i.e.\ as long as one is far from the phase-space
    boundary shown in (a).
    The rapidity is defined in the $q\bar q$ dipole centre-of-mass frame.
  }
  \label{fig:ME-1-parton}
\end{figure}

The essential properties of single-parton emission are illustrated in
Fig.~\ref{fig:ME-1-parton}.
There are three main elements to comment on regarding the above
analysis:
\begin{enumerate}
\item The effective single-emission matrix element in {\tt Pythia} and
  {\tt Dire}
  is correct in all singly-divergent regions of phase space, i.e.\
  both soft large-angle and hard-collinear, as well as soft-collinear.
  In {\tt Pythia} the invariance of the radiation pattern under boosts
  along the dipole direction is broken by running coupling effects: the
  same scale $\mu_R = v$ is used along the whole contour of constant
  $v$, even though, as one sees from
  Fig.~\ref{fig:ME-1-parton-const-v-contours}, that contour maps to a
  range of different physical $p_\perp$ values.
  This effect is expected to have consequences that are beyond NLL
  accuracy, because the region where $p_\perp$ differs substantially
  from $v$ comes with a finite weight only at large angle and a
  strongly suppressed weight in the anti-collinear region, cf.\
  Fig.~\ref{fig:ME-1-parton-dipole-partition-weights}.
  Accordingly we will not discuss it further in this article.
  
\item In both {\tt Pythia} and {\tt Dire}, the dipole is divided into two parts,
  one associated with the quark, the other with the anti-quark. That
  division occurs at zero-rapidity in the dipole rest frame, as is
  visible clearly in
  Fig.~\ref{fig:ME-1-parton-dipole-partition-weights}.
  While the sum adds up to one, the two elements of the partition
  behave differently for subsequent emissions, and the specific choice of
  partitioning can then have adverse consequences, as we shall see
  shortly.
  In particular it will affect subleading-$\nc$ LL terms, and the full
  set of NLL terms.

\item \label{enum:pythia-bad-LL} Within {\tt Pythia}, there is a suppressed but non-zero probability to
  have arbitrarily small kinematic transverse momentum, $|p_\perp|$,
  for any finite value of the evolution transverse-momentum variable
  $v\equiv p_{\perp,\evol}$ (i.e.\ the negative $\eta$ regions of
  Fig.~\ref{fig:ME-1-parton} for the $q\to q g$ splitting). 
  This too will have adverse consequences, at LL and leading-colour
  accuracy:
  if one asks what the probability is to have an event whose hardest
  emission is at some very small scale $p_{\perp,\text{cut}} \ll Q$,
  normally this implies a Sudakov suppression down to the scale
  $p_{\perp,\text{cut}}$.
  However in the {\tt Pythia} shower there is a second mechanism: full
  Sudakov suppression down to some intermediate scale
  $p_{\perp,\text{cut}} \ll p_{\perp,\evol} \ll Q$, together with
  creation of an emission with
  $|p_\perp| \sim p_{\perp,\text{cut}} \ll p_{\perp,\evol} $, which
  comes with a weight suppressed as in Eq.~(\ref{eq:py8-kt-eta-P}).
  These two mechanisms compete and for sufficiently small values of
  $p_{\perp,\text{cut}}$, the second one may dominate, which results
  in an overall degree of suppression of such configurations which no
  longer satisfies normal double-logarithmic scaling.
  In practice, the values of $\ln Q/|p_{\perp,\text{cut}}|$ at which
  this occurs are so large that they are unlikely to be of
  phenomenological interest.
  Accordingly we will not discuss this point any further here.
  Nevertheless, in future efforts to design showers, one should be
  aware that this kind of effect can arise.
\end{enumerate}

\subsection{Issues in two-emission case: double strong ordering}
\label{sec:double-emission-case-double}

We now consider double-emission configurations that probe the leading
singularity of the double-real squared amplitude, associated with
leading (double) logarithms.
The region of phase space that we will concentrate on for double
emission is that where the two emissions are both soft and collinear
to either of the hard partons and widely separated in rapidity from each
other, $|\eta_1 - \eta_2| \gg 1$.
In this limit, considering an initial $q\bar q$ dipole, the correct
double-emission probability has the very simple form
\begin{equation}
  \label{eq:two-emissions-correct}
  dP_{2} = \frac{C_F^2}{2!} \prod_{i=1,2} \left (
  \frac{2\as(p_{\perp,i}^2) }{\pi} \frac{dp_{\perp,i}}{p_{\perp,i}}
  d\eta_i \frac{d\phi_i}{2\pi}\right)\,. 
\end{equation}
where the $p_{\perp,i}$ and $\eta_i$ are defined with respect to the
$q$ and $\bar q$ directions.
Eq.~(\ref{eq:two-emissions-correct}) is valid even if
$p_{\perp,1} \sim p_{\perp,2}$.
The fundamental question that we ask is: do the parton showers
reproduce this?

To examine this question we will ignore the subtlety that the {\tt
  Pythia} kinematic $|p_\perp|$ can differ from the evolution
$|p_{\perp,\evol}|$, since this occurs with significant weight only in
a rapidity region of $\order{1}$, which corresponds to a soft emission
at wide angle.
Therefore the remainder of the discussion in this section will apply
equally well to the {\tt Pythia} and {\tt Dire} showers.

Let us follow the sequence of branchings that generates two emissions,
concentrating on the tree-level aspects:
\begin{itemize}
\item One starts with a $\bar q q$ dipole.
  Then a value $v_1$ of the ordering variable is chosen, together with
  an associated $z_1$, $\phi_1$.
  This leads to an emission of a first gluon $g_1$, with transverse
  momentum $p_{\perp,1}$ and rapidity $\eta_1$.
  One now has two dipoles: $\bar q g_1$ and $g_1 q$.
  
\item A value $v_2$ is chosen for the ordering variable, one selects
  one of the two current dipoles for branching, and within it one of
  its two ends.
  One then chooses the $z_2$ and $\phi_2$ splitting variables and
  generates a second gluon $g_2$.
  To understand the effective matrix element one should consider the
  sum over all four resulting situations.
\end{itemize}
Let us first consider the four cases in the situation where
$v_2 \ll v_1$, which has the simplification that for the phase-space
regions of interest, $|p_{\perp,i}| = v_i$.
For convenience we will henceforth take $Q=1$.
\begin{enumerate}
\item $\bar q \to \bar q g_2$ splitting of the $\bar q g_1$ dipole,
  for which we adopt the shorthand $\bar q[g_1] \to \bar q g_2 [g_1]$,
  putting the spectator parton of the dipole in square brackets.
  The region of phase-space that gets filled and the associated
  splitting weight are as follows (we remind the reader that emissions
  collinear to $\bar q$ have negative $\eta$):
  \begin{equation}
    \label{eq:double-region1}
    \ln v_2 \; \ll \; \eta_2 \;\ll\;
    \frac12\left(\eta_1 + \ln v_1\right)
    \quad \to \quad
    d\cP_2 = C_F\frac{2\as(|p_{\perp,2}^2|)}{\pi} d\eta_2 \frac{dp_{\perp,2}}{p_{\perp,2}}\,.
  \end{equation}
  The notation $a\ll b$ for logarithmic variables like rapidities
  should be understood as meaning that $e^{a-b}$ is small.
  The left-hand bound on $\eta_2$ corresponds to the maximum allowed
  (negative) rapidity along the anti-quark direction, i.e.\ the
  hard-collinear limit for radiation from the $\bar q$.
  The right-hand bound is determined as the point of zero rapidity of
  the $\bar q g_1$ dipole in its own rest frame, but translated into
  the original $\bar q q$ rest frame.
  Both bounds are given to within corrections of $\order{1}$, which
  are irrelevant for the purpose of our discussion, since they
  generate subleading logarithmic corrections.
  
\item $g_1[\bar q] \to g_1 g_2[\bar q]$:
  part of this branching is collinear to gluon $1$ (in the
  centre-of-mass frame of the hard scattering), and we ignore that
  part for now.
  The remainder is collinear either to the $\bar q$ or $q$, covering a
  rapidity region
  \begin{equation}
    \label{eq:double-region2}
    \frac12\left(\eta_1 + \ln v_1\right) \; \ll \; \eta_2 \;\ll\;
    \eta_1
    \quad \to \quad
    d\cP_2 = \frac{C_A}{2}
        \frac{2\as(|p_{\perp,2}^2|)}{\pi} d\eta_2 \frac{dp_{\perp,2}}{p_{\perp,2}}\,.
  \end{equation}
  This splitting weight $d\cP_2$ here is identical to that in Eq.~(\ref{eq:double-region1}), except for the
  replacement of $C_F \to C_A/2$.
  This will be a source of problems: one thinks of the
  $g_1 \to g_1 g_2$ splitting as being the emission of a gluon from a
  gluon hence the $C_A/2$ colour factor.
  However splitting the $\bar q g_1$ dipole into two equal parts in
  its rest frame causes some part of the radiation assigned to the
  gluonic part to be in a phase space region where it is closer in
  angle to the $\bar q$ or $q$ than it is to the gluon.
  In that region, the $C_A/2$ colour factor is wrong.
  
\item $g_1[q] \to g_1 g_2[q]$,
  which is
  analogous to Eq.~(\ref{eq:double-region2}) but with $\bar q \to
  q$,
  \begin{equation}
    \label{eq:double-region3}
    \eta_1
    \; \ll \; \eta_2 \;\ll\;
    \frac12\left(\eta_1 + \ln \frac{1}{v_1}\right)
    \quad \to \quad
    d\cP_2 = \frac{C_A}{2}
    \frac{2\as(|p_{\perp,2}^2|)}{\pi} d\eta_2 \frac{dp_{\perp,2}}{p_{\perp,2}}\,,
  \end{equation}
  and again with the erroneous $C_A/2$ colour factor.

\item $q[g_1] \to q g_2[g_1]$
  which is
  analogous to Eq.~(\ref{eq:double-region1}),
    \begin{equation}
    \label{eq:double-region4}
    \frac12\left(\eta_1 + \ln \frac{1}{v_1}\right)
    \; \ll \; \eta_2 \;\ll\;
    \ln \frac1{v_2}
    \quad \to \quad
    d\cP_2 = C_F\frac{2\as(|p_{\perp,2}^2|)}{\pi} d\eta_2 \frac{dp_{\perp,2}}{p_{\perp,2}}\,.
  \end{equation}
\end{enumerate}%
\begin{figure}
  \centering
  \begin{subfigure}[t]{0.49\textwidth}
    \includegraphics[width=\textwidth]{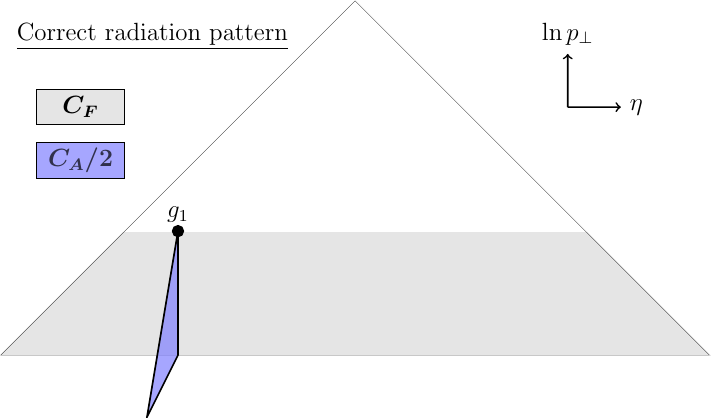}
    \caption{}
    \label{fig:lund-subleading-nc-correct}
  \end{subfigure}
  \hfill%
  \begin{subfigure}[t]{0.49\textwidth}
    \includegraphics[width=\textwidth]{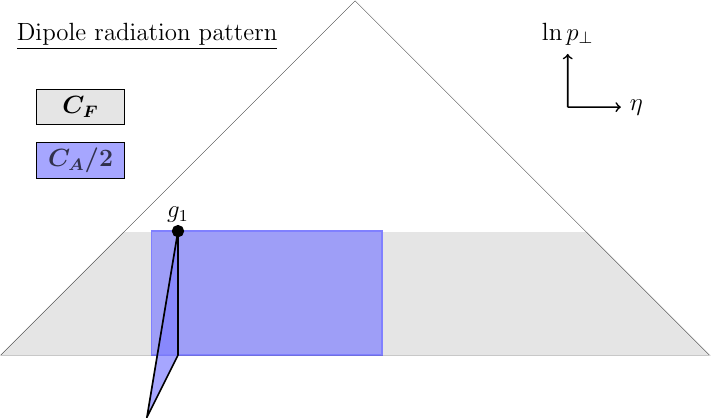}
    \caption{}
    \label{fig:lund-subleading-nc-dipole}
  \end{subfigure}
  \caption{Lund-diagram illustrations of the subleading-$N_C$ issue in
    the showers that we consider.
    As a starting point we take a right (left)-moving quark
    (anti-quark), and gluon $g_1$ emitted at the $\eta{-}\ln p_\perp$
    coordinate shown in the big (``primary'') triangle.
    The phase-space for emission of a further gluon from the $q g_1$
    dipole corresponds to the shaded 
    area to the right of $g_1$ on the primary triangle, and the right-hand
    face of the ``leaf'' that comes out of the plane;
    analogously the phase-space for emission from the $\bar q g_1$
    dipole corresponds to the shaded area of the primary triangle to the
    left of $g_1$ and to the left-hand face of the leaf.
    The colour factor associated with the phase-space region is
    indicated by the colour of the shading: 
    grey denotes $C_F$, while blue denotes $C_A/2$.
    The left-hand diagram shows the correct pattern, the right-hand
    diagram shows the outcome of the \pythia and \dire showers.  }
  \label{fig:lund-subleading-nc}
\end{figure}%
The main message to retain from this analysis is that there is a region
that has both soft and collinear enhancements, for each of the two
emissions, where instead of a $C_F^2$ colour factor, one obtains a
$C_F C_A/2$ colour factor, i.e.\ an incorrect subleading $\nc$ term.
This is illustrated in the Lund diagram of
Fig.~\ref{fig:lund-subleading-nc}: panel
(\subref{fig:lund-subleading-nc-correct}) shows the correct assignment
of colour factors across phase-space for radiation below the scale of
$g_1$.
The coloured ``leaf'' that comes out of the plane represents the
additional phase-space that opens up following emission of $g_1$, with
a $C_A/2$ colour factor associated with each of its two faces.
The restriction of the phase-space to that region is a consequence of
angular ordering, as discussed for example some time ago in
Ref.~\cite{Gustafson:1992uh}. 
Panel (\subref{fig:lund-subleading-nc-dipole}) shows the assignment
that is effectively made in the case of the \pythia and \dire showers, with the
coloured area ($C_A/2$) now extending into the primary Lund
triangle.\footnote{Note that since we start with a $q\bar q$ system,
  the primary plane emits only from the front face. For an initial
  $gg$ system, one might instead choose to represent emissions from
  both the front and rear faces, reflecting the presence of two
  $C_A/2$ dipoles.}
Since regions with simultaneous soft and collinear enhancements (i.e.\
extended areas in the Lund diagram) tend to be associated with leading
double logarithms in distributions of common observables, one may
expect that this issue with subleading $\nc$ terms will also affect
those double logarithms.
We will investigate this in section~\ref{sec:NLO-LL-subl-n_c}.

We should note that issues with the attribution of colour factors
beyond leading $N_C$ in dipole showers have been highlighted in a
range of previous work, e.g.\
Refs.~\cite{Friberg:1996xc,Giele:2011cb,Nagy:2012bt,Hartgring:2013jma}. 
Our analysis in this subsection is close in particular to that of
Ref.~\cite{Friberg:1996xc}. 
We also note that approaches to obtain the correct subleading colour
factor for at least the main soft-collinear divergences have existed
for some time.
The classification that is implied by angular ordering (see also
Ref.~\cite{Gustafson:1992uh}) provides a guide in this direction, as
was articulated for a dipole shower in Ref.~\cite{Friberg:1996xc} and
found to be relevant for particle multiplicities at LHC
energies~\cite{Eden:1998ig}.
Another proposal is that of Ref.~\cite{Giele:2011cb}.


\begin{figure}[t!]
    \centering
    \begin{subfigure}[t]{0.53\textwidth}
        \centering
        \includegraphics[width=\textwidth]{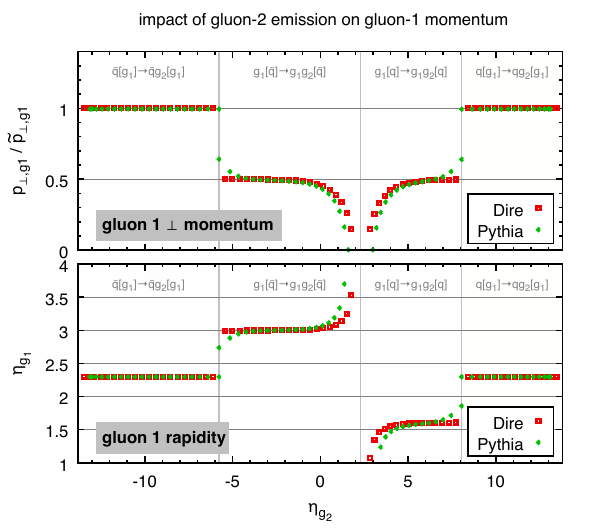}
        \caption{}
        \label{fig:double-line-test}
    \end{subfigure}%
    \hfill
    \begin{subfigure}[t]{0.43\textwidth}
        \centering
        \includegraphics[width=\textwidth]{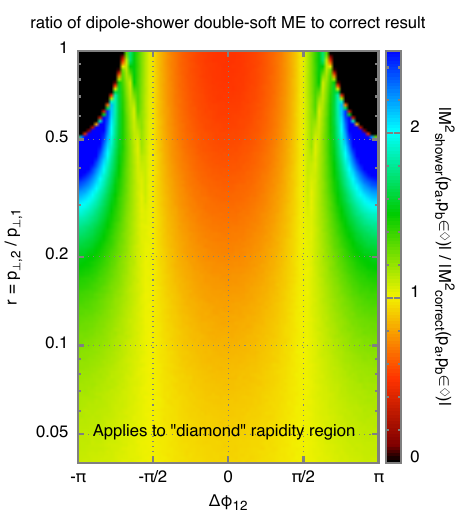}
        \caption{}
        \label{fig:double-soft-ME-ratio}
    \end{subfigure}
    \caption{(a) Illustration of the modification of the transverse momentum
    (upper panel) and rapidity (lower panel) of
    gluon $1$ after emission of gluon $2$,
    shown as a function of the rapidity of gluon $2$.
    Prior to emission of gluon 2, gluon $1$ originally has a rapidity
    $\eta_{g_1} \simeq 2.3$ and transverse momentum
    $\ptilde_{\perp,g_1} = v_1 = 10^{-6} Q$ ($v_1 = 10^{-6 }Q$ and
    $1-z_1 = 10^{-5}$). Gluon $2$ has $v_2 = \frac12 v_1$ and is
    emitted parallel in azimuth to gluon $1$.
    To help guide the eye, four regions of gluon 2 rapidity are
    labelled according to the identity of the parton that branches and
    that of the spectator.
    The results have been obtained using a numerical implementation of
    the kinematic maps of section~\ref{sec:shower-summary}.
    The transverse momentum shifts in (a) can be reinterpreted in
    terms of the effect they have on the effective matrix element for
    double-soft emission.
    Plot (b) shows the ratio of this effective
    matrix element to the true one, as a function of the azimuthal angle
    between the two emissions and their transverse-momentum ratio (in
    a specific ``diamond'' region of widely separated rapidities, cf.\
    Appendix~\ref{sec:app:double-soft-ME}).
    For simplicity, the matrix-element ratio is given in the
    large-$\nc$ limit.}
\end{figure}

\subsection{Issues in two-emission case: single strong ordering}
\label{sec:double-emission-case-single}

Now we turn to the case where $v_2$ is only moderately smaller than
$v_1$.
Again one may consider the four cases listed in section~\ref{sec:double-emission-case-double}, and in each case
we will determine the kinematics of the four final-state partons.
It is easiest to first illustrate what happens with reference to
Fig.~\ref{fig:double-line-test}.
Here we have generated a sequence of two emissions, $g_1$ and $g_2$,
with $v_2 = v_1/2$,
and we study how the momentum of $g_1$ is modified after emission of
$g_2$.
Using $\boldsymbol{\ptilde}_{\perp,g_1}$ and $\etatilde_{g_1}$ ($\boldsymbol{p}_{\perp,g_1}$
and $\eta_{g_1}$) to denote the 2d-vector transverse momentum and
rapidity respectively of gluon $g_1$ before (after) emission of $g_2$,
the figure illustrates the following pattern of modifications:
\begin{equation}
  \label{eq:recoil-impact}
  \begin{array}{lllll}
    1.\quad \bar q[g_1] \to \bar q g_2 [g_1] :
    &\quad& \boldsymbol{p}_{\perp,g_1} = \boldsymbol{\ptilde}_{\perp,g_1}\,,
    &\quad& \eta_{g_1} = \etatilde_{g_1}\,,
    \\
    2.\quad g_1[\bar q] \to g_1 g_2 [\bar q] :
    &\quad& \boldsymbol{p}_{\perp,g_1} =
       \boldsymbol{\ptilde}_{\perp,g_1} - \boldsymbol{p}_{\perp,g_2}
       \,,
    &\quad& \eta_{g_1} = \etatilde_{g_1} - \ln\frac{|\boldsymbol{p}_{\perp,g_1}|}{|\boldsymbol{\ptilde}_{\perp,g_1}|}\,,
    \\
    3.\quad g_1[q] \to g_1 g_2 [q] :
    &\quad& \boldsymbol{p}_{\perp,g_1} =
       \boldsymbol{\ptilde}_{\perp,g_1} - \boldsymbol{p}_{\perp,g_2}
       \,,
    &\quad& \eta_{g_1} = \etatilde_{g_1} + \ln\frac{|\boldsymbol{p}_{\perp,g_1}|}{|\boldsymbol{\ptilde}_{\perp,g_1}|}\,,
    \\
    4.\quad q[g_1] \to q g_2 [g_1] :
    &\quad& \boldsymbol{p}_{\perp,g_1} = \boldsymbol{\ptilde}_{\perp,g_1}\,,
    &\quad& \eta_{g_1} = \etatilde_{g_1}
  \end{array}
\end{equation}
In regions $1$ and $4$, gluon $1$ remains essentially unaffected by
the emission of $2$ (the transverse recoils are absorbed by the
quark). 
This is correct, because in the exact matrix element, soft gluons that
are widely separated in rapidity are independent of each other.
In regions $2$ and $3$, where $g_2$ is at relatively central
rapidities, the situation is different: $g_1$ acquires a transverse
recoil to balance the transverse momentum of $g_2$: this causes the
$p_{\perp,g_1} / \ptilde_{\perp,g_1}$ to be equal to $\frac12$ in the
corresponding regions of Fig.~\ref{fig:double-line-test}. 
There is also a corresponding modification of the rapidity of $g_1$
and its sign and magnitude can be worked out by noting that the dipole
mass must be conserved despite the modification of the transverse
momentum of $g_1$, i.e.\ by imposing that
$p_{\perp,g_1} e^{\pm\eta_{g_1}} = \ptilde_{\perp,g_1}
e^{\pm\etatilde_{g_1}}$, where the choice of sign depends on the
specific configuration.

These modifications of the transverse momentum and rapidity of gluon 1
after emission of a subsequent gluon 2 are a cause for concern.
This is most easily seen by working out the effective 
splitting weight for the emission of two soft gluons in regions $2$ and
$3$.
We concentrate on a specific ``diamond''
rapidity region, which has single-logarithmic rapidity enhancements
for each of the gluons, and whose size is $1/3$ of the total double
rapidity phase-space.
The details and analysis are given in
Appendix~\ref{sec:app:double-soft-ME}, and we concentrate here on the
results.
The result for the ratio of the effective matrix element to the
correct one, Eq.~(\ref{eq:two-emissions-correct}), is shown in
Fig.~\ref{fig:double-soft-ME-ratio} as a function of the azimuthal
angle between the two emissions and their transverse-momentum ratio.
The figure reveals some unwanted features.
These include the empty zones for
$p_{\perp,2}/p_{\perp,1} \gtrsim \frac12$ and
$|\Delta \phi_{12}| \gtrsim 2\pi/3$ and the strong enhancement in a
similar azimuthal region for
$\frac14 \lesssim p_{\perp,2}/p_{\perp,1} \lesssim \frac12$.
There is also depletion and enhancement in other areas of the plot.
Only for rather small values of $p_{\perp,2}/p_{\perp,1}$ does the
effective shower matrix element tend to the correct result.

Some of the features of Fig.~\ref{fig:double-soft-ME-ratio} are
straightforward to understand qualitatively.
Consider, for example, the case when the second gluon is emitted
back-to-back with respect to the first, $\Delta \phi_{12}=\pi$ and
with a $p_{\perp,2}$ that is a fraction $\widetilde r$ of the first
emission's original $\ptilde_{\perp,1}$.
The first emission's transverse momentum gets increased by a factor
of $1+\widetilde r$, so that the new ratio of transverse momenta
becomes $r = \widetilde r/(1+\widetilde r).$
Since $\widetilde r \le 1$, the final ratio $r$ is bounded to be
less than $r \le 1/2$.
This generates the dead zone for $r > \frac12$ and the strong
enhancement just below.

As we will see below, the underlying recoil issue that leads to the
incorrect double-soft tree-level matrix element will also cause many
common observables, e.g.\ $e^+e^-$ event shapes, to have incorrect NLL
(leading-$\nc$) terms in distributions as evaluated with parton
showers.

The question of recoil in showers with dipole-local recoil was first
raised long ago~\cite{Andersson:1991he}.
That analysis compared the effective shower matrix element to the
full double-emission matrix element in $e^+e^-$ collisions.
In particular it highlighted the dead-zone problem that is visible
near $\Delta\phi_{12} = \pm \pi$ and $r\simeq 1$ in our
Fig.~\ref{fig:double-soft-ME-ratio}.
However, it differs from our analysis in that it did not take the
formal soft and collinear limits, and as such could also not extend to
the logarithmically-relevant limit of widely separated rapidities for
the two emissions.

More recently, Nagy and Soper (e.g.\
Refs.~\cite{Nagy:2009vg,Nagy:2014nqa}) and also Gieseke and
Pl\"atzer~\cite{Platzer:2009jq} (journal version) have highlighted
issues to do with recoil in the context of the Drell-Yan (DY)
transverse momentum distribution.
The final-state recoil issue that we have discussed here is intimately
connected with those recoil issues in the initial state.
The novelty of our finding is that this type of issue applies not just
to a specific observable, and not just to processes with initial-state
hadrons, but to the full pattern of soft gluon radiation in
essentially any process.
Aspects introduced by Nagy and Soper, in particular the combination of
a time-like ordering variable and a modified (global) recoil scheme,
can, we believe, be critical ingredients in addressing the recoil
problem that we are discussing.
However the Nagy-Soper shower prescriptions are significantly more
complicated than the \pythia or \dire showers and include a number of
variants (e.g.\ both $p_\perp$ and time ordering).
As we have seen in the \pythia single-emission case, e.g.\ issue~\ref{enum:pythia-bad-LL}  in
section~\ref{sec:single-emission-case}, subtleties can arise in almost
any aspect of a shower, and a conclusion should only be drawn from a
full, detailed analysis of a specific shower prescription.

A further point to be aware of is that any analysis of a proposed
solution needs to go beyond the two-emission case.
In particular, qualitatively new recoil-related issues can arise
starting from the third emission.
For example, the solution proposed in Ref.~\cite{Andersson:1991he} and
adopted in the \ariadne program~\cite{Lonnblad:1992tz}, is to assign
recoil for a $qg$ dipole to the quark.
While we believe this to be adequate at second order, the 
recoil issue reappears at 3rd order for emission from a $gg$
dipole, and is not, we believe, addressed by the solution of
Ref.~\cite{Andersson:1991he}.

\section{Logarithmic analysis at second order}
\label{sec:fixed-order}

While section~\ref{sec:singularity-structure} illustrated physical
shortcomings of two widespread showers, the key question that
remains to answer is that of the consequences of those shortcomings.
Insofar as a parton shower is supposed to provide resummation of
logarithms, the natural way of examining those consequences is in
terms of the impact on the logarithmic accuracy of the shower
predictions for various classes of observable.

We will discuss the logarithmic accuracy in the context of event-shape
variables, which have been widely studied and are well understood from
the point of analytic and semi-numerical resummation.
Let $V$ be some event shape variable, a function of all the momenta in
an event.
The quantity we will study is $\Sigma(L)$, the probability that the
event shape has a value smaller than $e^{-L}$.
For most event-shape variables the structure of $\Sigma(L)$ is of the
form
\begin{subequations}
  \begin{align}
    \label{eq:form-lnSigma}
    \Sigma(L) &= \exp \left[L g_1(\as L) + g_2(\as L) + \as g_3(\as L) +
                \cdots \right] + \order{\as e^{-L}}\,,
    \\
              &= \exp \left[\sum_{m=1}^\infty \sum_{n=0}^{m+1} G_{mn} \as^m L^n \right] + \order{\as e^{-L}}\,,
  \end{align}
\end{subequations}
where we emphasise the logarithmically enhanced part of the result.
In our default counting of logarithmic accuracy, the $\ln \Sigma$ counting,
the $g_1(\as L)$ function contains the LL terms in $\ln \Sigma$,
$\as^n L^{n+1}$;
the $g_2(\as L)$ function contains the NLL terms, $\as^n L^n$; and so
forth.
Results up to NLL can be obtained for arbitrary global event shape
type observables using the CAESAR formalism~\cite{Banfi:2004yd}.
The formalism classifies observables in terms of two parameters, $a$
and $b$ according to their dependence on the momentum of a single
soft-collinear emission
\begin{equation}
  \label{eq:caesar-scaling}
  V(p,\{\text{Born momenta}\})
  \,\propto \,
  p_\perp^a e^{-|\eta_p| b}\,,
\end{equation}
a classification that we will refer to below.
The use of $\as^{\text{CMW}}$, cf.\
Eqs.~(\ref{eq:Dire-Ingredients-q-qg-branching-prob})--(\ref{eq:3}) is
sometimes held to be sufficient to reproduce NLL accuracy.
While it is a necessary ingredient, together with two-loop running of
the coupling, on their own these elements are not sufficient.
Event shape observables are interesting to consider, with a view to
future parton shower developments, because in many cases their
resummation is known to NNLL or even higher
accuracy~\cite{Becher:2008cf,Monni:2011gb,Hoang:2014wka,Becher:2012qc,Banfi:2014sua,Banfi:2016zlc}.

There is also an alternative, $\Sigma$ counting, reflecting a
structure
\begin{equation}
  \label{eq:form-Sigma}
  \Sigma(L) = \sum_{m=0}^\infty \sum_{n=0}^{2m}  h_{m,n}\, \as^m L^n
   + \order{\as e^{-L}}
  \,.
\end{equation}
In this counting, LL$_\Sigma$ terms correspond to the $\as^n L^{2n}$
contributions, NLL$_\Sigma$ to $\as^n L^{2n-1}$ and so forth.
%
%
There are some instructive observables for which only
Eq.~(\ref{eq:form-Sigma}) can be used.
These include $n$-jet rates in $e^+e^-$ for $n\ge 3$ with
Cambridge~\cite{Dokshitzer:1997in} and Durham~\cite{Catani:1991hj} jet
clustering.\footnote{And also many non recursively infrared and
  collinear safe observables~\cite{Banfi:2004yd}, such as the JADE jet
  rates~\cite{Brown:1990nm}.} 
The $n$-jet rates are interesting also because the same physics that
is required for their correctness enters into the calculation of the
scaling of hadron multiplicities with energy~\cite{Marchesini:1983bm}.

\subsection{Subleading-$\nc$ $\as^2 L^4$ terms}
\label{sec:NLO-LL-subl-n_c}

Observables will break up into two basic classes from the point of
view of subleading-$\nc$ $\as^2 L^4$ terms: those with $b\ne 0$ and
those with $b=0$ in Eq.~(\ref{eq:caesar-scaling}).
Let us start with the example of the thrust~\cite{Farhi:1977sg}, i.e.\
a case with $a=1$, $b=1$.
The LL$_\Sigma$ result for the thrust is
\begin{equation}
  \label{eq:thrust-DL}
  \Sigma(L)
  \;=\;
  \exp \left( - \frac{\abar L^2}{2} \right)
  \;=\;
  1 \,-\, \frac{\abar L^2}{2} \,+\, \frac{\abar^2 L^4}{8} + \order{\abar^3}\,,
\end{equation}
where
\begin{equation}
  \label{eq:abar}
  \abar = \frac{2 \as C_F}{\pi}\,.
\end{equation}
\begin{figure}
  \centering
  \includegraphics[width=0.6\textwidth]{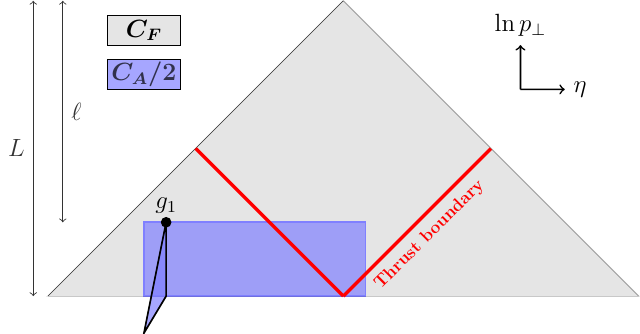}
  \caption{Lund diagram to help illustrate subleading-$\nc$ issue for the
    Thrust. See text for further details.}
  \label{fig:subleading-NC-thrust}
\end{figure}%
To illustrate how the colour-factor issue of
Eqs.~(\ref{eq:double-region2}) and (\ref{eq:double-region3}) impacts the
double-logarithmic structure, it is useful to consider
Fig.~\ref{fig:subleading-NC-thrust}.
For a given constraint on the thrust, one has the thick red boundary:
emissions above that boundary are vetoed.
However there can be emissions below that boundary, e.g.\ emission
$g_1$, which modify the colour factor for subsequent emissions (and
their associated virtual corrections) with a lower value of the
ordering variable $v$, but that are above the thrust boundary: instead of being associated with the correct
$C_F$ factor, they have a factor $C_A/2$.
The region where this occurs is shown in blue, and corresponds to the
rapidity regions of
Eqs.~(\ref{eq:double-region2}) and (\ref{eq:double-region3}), cf.\
also Fig.~\ref{fig:lund-subleading-nc-dipole}.
The second-order, LL$_\Sigma$ issue that arises because of this can be
evaluated by considering the area of the blue region that is above the
thrust boundary, integrating also over the phase space for real emission
$1$,
\begin{subequations}
  \label{eq:thrust-DL-subleading-issue}
  \begin{align}
    \delta \Sigma(L)
    &= -2\abar^2 \int_{L/2}^L d\ell_1
      \int_{-L + \ell_1}^{-\ell_1} d\eta_1
      \int_{\ell_1}^L d\ell_2
      \int_{\frac12(\eta_1 -\ell_1)}^{\frac12(\eta_1 +\ell_1)} \!d\eta_2\,
      \Theta(|\eta_2| < L - \ell_2)
    \left(\frac{C_A}{2C_F} - 1 \right),
    \\
    &= -\frac{1}{64} \abar^2 L^4 \left(\frac{C_A}{2C_F} - 1 \right)\,,
  \end{align}
\end{subequations}
where one should recall that $L$ is positive, we have introduced
$\ell_i = \ln 1/v_i$, $\abar$ includes a $C_F$ colour factor, cf.\
Eq.~(\ref{eq:abar}), and there is an overall factor of two associated
with the possibility of $\eta_1$ being either negative or positive
(the integral includes only the negative case).
This correction is double logarithmic.
However it is $\nc$ suppressed, by a factor
\begin{equation}
  \label{eq:def-c}
  c \equiv \left(\frac{C_A}{2C_F} - 1 \right)
  = \frac{1}{\nc^2 - 1} = \frac18 \,.
\end{equation}

A double-logarithmic (LL$_\Sigma$) $\nc$-suppressed effect of this kind is present
for any event-shape like observable with $b \neq 0$ in
Eq.~(\ref{eq:caesar-scaling}). 
The fact that terms $\as^n L^{2n}$ are modified for $n \ge 2$ means
that the result from the parton shower does not properly exponentiate
beyond leading colour, i.e.\ cannot be written in the form of
Eq.~(\ref{eq:form-lnSigma}). 

In contrast, for observables with $b=0$, for example the jet
broadening, there is no such effect at double-logarithmic accuracy.
This is because the boundary associated with a limit on the value of a
$b=0$ observable corresponds to a horizontal line in the Lund plane.
As a result the only region in which it matters that an emission $1$
modifies the colour for a subsequent emission $2$ is when both
emissions have a commensurate transverse momentum.
This removes a number of logarithms and induces only a
NLL$_{\ln \Sigma}$ type $\nc$-suppressed effect.

We expect similar issues of a wrong subleading-colour coefficient
for the double logarithms in the $e^+e^-$ 4-jet (and higher) rates
with the $k_t$ algorithm~\cite{Catani:1991hj}.
Note that these are somewhat different from the (leading-colour)
issues discussed in Ref.~\cite{Webber:2010vz} for other, non-dipole
classes of $p_\perp$ ordered shower.
%

\subsection{Leading-$\nc$ $\as^2 L^2$ terms}
\label{sec:NLO-NLL}

If we work in the leading-$\nc$ limit, $C_A = 2C_F$, then the impact
of the incorrect shower mappings in regions $2$ and $3$ of
Eq.~(\ref{eq:recoil-impact}) can be written as follows (recall that we
are using $Q=1$)
\begin{multline}
  \label{eq:NLL-as2-master}
  \delta \Sigma(L) = 
  \abar^2
  \int_0^1 \frac{dv_1}{v_1}
  \int_{\ln v_1}^{\ln 1/v_1} d\eta_1
  \int_0^{v_1} \frac{dv_2}{v_2}
  \int_{\frac12(\eta_1 + \ln v_1)}^{\frac12(\eta_1 + \ln 1/v_1)} d\eta_2
  \int_0^{2\pi}
  \frac{d\phi_{12}}{2\pi}
  \times
  \\
  \times
  \left[\Theta\big(e^{-L} - V(p_1^\text{shower},p_2)\big) 
    - \Theta\left(e^{-L} - V(p_1^\text{correct},p_2)\right)
  \right],
\end{multline}
where we examine the difference between the double-real contribution
with a ``shower'' mapping and a correct mapping.
``Correct'' means any mapping that leaves the transverse momentum and
rapidity of $p_1$ unchanged for $|\eta_1 - \eta_2| \gg 1$ and so
reproduces the Abelian limit.
Eq.~(\ref{eq:NLL-as2-master}) holds in the soft and collinear limit
and for compactness in the arguments of $V$ we omit the momenta of the
(hard) quark and anti-quark, keeping in mind that in any practical
shower implementation they must of course be included.
We do not need to consider virtual corrections because from the
kinematic point of view any configuration with fewer than two emissions
has the correct leading-$\nc$ distribution of emitted partons and
hence the virtual contribution cancels in the difference between
``correct'' and ``shower'' cases.
We have omitted the $\phi_{1}$ azimuthal integral, and written the
$\phi_2$ integral in terms of $\phi_{12} = \phi_2 - \phi_1$.
We work in a fixed-coupling limit, for simplicity.

To obtain a concrete result from Eq.~(\ref{eq:NLL-as2-master}) we
first consider the $2$-jet rate in the Cambridge $e^+e^-$ jet
algorithm~\cite{Dokshitzer:1997in}, which is akin to calculating the
distribution of $V$ defined as $\sqrt{y_\text{cut}}$.\footnote{It is
  also closely related to jet-veto survival factors at hadron
  colliders.}
The Cambridge algorithm has the simple property that
$V(\{p_i\}) = \max_i \{p_{\perp,i}\}$ for soft collinear emissions
that are widely separated in rapidity.
This allows us to write
\begin{equation}
  \label{eq:4}
  V(p_1^\text{correct},p_2) = v_1\,\qquad
  V(p_1^\text{shower},p_2) = \max\left(v_2,\, \sqrt{v_1^2 + v_2^2 - 2v_1 v_2 \cos
  \phi_{12}}\right).
\end{equation}
The absence of dependence on the particle rapidities makes it
straightforward to evaluate the $\eta_1$ and $\eta_2$ integrals, and
it will also be convenient to introduce $\zeta = v_2/v_1$.
We can then write 
\begin{multline}
  \label{eq:NLL-as2-cam}
  \delta \Sigma^\text{cam}(L) = 
  \abar^2 \int_0^1 \frac{dv_1}{v_1}  2 \ln^2 \frac1{v_1}
  \int_0^1 \frac{d\zeta}{\zeta}
  \int_0^{2\pi}
  \frac{d\phi_{12}}{2\pi}
  \times
  \\
  \times
  \left[
    \Theta\left(e^{-L} - v_1 \max\left(\zeta,\sqrt{1 + \zeta^2 -
          2\zeta \cos\phi_{12}}\right)\right) 
    - \Theta\left(e^{-L} - v_1 \right)
  \right].
\end{multline}
This reduces to
\begin{subequations}
  \label{eq:NLL-as2-cam-res}
  \begin{align}
    \delta \Sigma^\text{cam}(L)
    &= 
      2 \abar^2 L^2   \int_0^1 \frac{d\zeta}{\zeta}
      \int_0^{2\pi}
      \frac{d\phi_{12}}{2\pi}
      \ln \frac{1}{\max\left(\zeta, \sqrt{1 + \zeta^2 - 2\zeta
      \cos\phi_{12}}\right)}
      + \order{\abar^2 L}\,,
    \\
    &=
      -0.18277 \, \abar^2 L^2  + \order{\abar^2 L}\,.
  \end{align}
\end{subequations}
This demonstrates the presence of a NLL deficiency that starts at order
$\as^2$.

Another simple observable is the fractional moment of the
energy-energy correlation, $\text{FC}_1$, defined in appendix~I.2 of
Ref.~\cite{Banfi:2004yd}, which reduces to $V(\{p_i\}) = \sum_i
p_{\perp,i}$ (i.e.\ a scalar sum) in the soft-collinear limit, giving
\begin{subequations}
  \label{eq:NLL-as2-FC1-res}
  \begin{align}
    \delta \Sigma^{\text{FC}_1}(L)
    &= 
      2 \abar^2 L^2   \int_0^1 \frac{d\zeta}{\zeta}
      \int_0^{2\pi}
      \frac{d\phi_{12}}{2\pi}
      \ln \frac{1 + \zeta}{ \sqrt{1 + \zeta^2 - 2\zeta
      \cos\phi_{12}} + \zeta}
      + \order{\abar^2 L}\,,
    \\
    &=
      -0.066934 \, \abar^2 L^2  + \order{\abar^2 L}\,.
  \end{align}
\end{subequations}
The numerical coefficients in
Eqs.~(\ref{eq:NLL-as2-cam-res},\ref{eq:NLL-as2-FC1-res}) are not
particularly large.
Nevertheless they can be relevant, especially from the perspective of
trying to obtain accurate parton showers for the LHC.
Consider high-$p_\perp$ jets of a few TeV, where one might probe the
substructure using shapes such $N$-subjettiness
ratios~\cite{Thaler:2010tr}.
If one is sensitive to radiation at the $5\GeV$ scale where
$\as \simeq 0.2$, one finds $\abar L \simeq 1$ and so
Eq.~(\ref{eq:NLL-as2-cam-res}) would point to effects of the order of
$20\%$.
Another point of comparison is to the effect of the CMW correction
(cf.\ Eq.~(\ref{eq:3})),
which for both these observables reads $ \abar L^2 \times (\alpha_s/2\pi)
K =  \abar^2 L^2 K/(4 C_F) \simeq 0.65\, \abar^2 L^2$.
In a context where groups are seeking to develop showers with
higher-accuracy splitting
kernels~\cite{Jadach:2010aa,Li:2016yez,Hoche:2017hno,Hoche:2017iem,Dulat:2018vuy},
phenomenologically such an effect should not be neglected.

Interestingly there are also observables for which the $\as^2 L^2$
coefficient is zero.
Perhaps the most notable is anything that relates to a vector sum over
the emissions' transverse momenta. 
Keeping in mind that
$\boldsymbol{p}_{\perp,1}^\text{shower} =
\boldsymbol{\ptilde}_{\perp,1} - \boldsymbol{p}_{\perp,2}$, the shower
vector sum,
$\boldsymbol{p}_{\perp,1}^\text{shower} + \boldsymbol{p}_{\perp,2}$ is
simply equal to $|\boldsymbol{\ptilde}_{\perp,1}|=v_1$, while the
correct result is $v_1 \sqrt{1 + \zeta^2 + 2\zeta \cos\phi_{12}}$.
Since the following integral vanishes
\begin{equation}
  \label{eq:vecpt-vanishes}
  \int_0^{2\pi}
  \frac{d\phi_{12}}{2\pi}
  \ln \left(1 + \zeta^2 + 2\zeta \cos\phi_{12}\right)
  = 0\,\qquad\qquad \text{for } 0<\zeta<1\,,
\end{equation}
there will be no $\as^2 L^2$ error for any observable that reduces to
such a vector sum.
Vector-sum type observables and deficiencies of transverse-momentum
showers with dipole-local recoil schemes have seen some discussion for
initial-state showering.
Nagy and Soper~\cite{Nagy:2009vg}  noted that it could affect
logarithmic accuracy, though we are not aware of a specific statement
detailing what accuracy would be affected.
An explicit study of local versus global recoil schemes in Appendix~C
of the {\tt Dire} paper~\cite{Hoche:2015sya} suggested that the numerical
impact is small.
This would not be surprising if our analysis here carries over to the
initial-state case and implies a zero $\as^2 L^2$ coefficient there
too.
Note that an $\as^3L^3$ study that we have carried out shows that the
zero is not an all-order property.\footnote{ For example, for an
  observable that reduces to the vector sum of the transverse momenta
  of all soft emissions in the two hemispheres of an $e^+e^-$ event,
  there is an erroneous NLL contribution to $\Sigma(L)$ which starts
  with a term $\simeq -0.250 \abar^3 L^3$.
  Such an observable is similar to the transverse momentum of the
  $Z$-boson in hadron--hadron collisions.
  For $e^+e^-$ collisions we are not aware that such an observable has
  ever been explicitly studied, however we believe it should be
  possible to construct it, for example starting from the observation
  in Appendix~I.1 of Ref.~\cite{Banfi:2004yd}, that certain
  Berger-Kucs-Sterman angularities~\cite{Berger:2003iw} effectively reduce to hemisphere
  vector sums.}

We have analysed two further observables that are somewhat more
involved: the total jet broadening~\cite{Catani:1992jc} has a non-zero
$\as^2 L^2$ coefficient, while the thrust is zero at $\as^2 L^2$ but
not zero at $\as^3 L^3$ (an all-order analysis reveals further
subtleties, however). 
The situation is summarised in Table~\ref{tab:fixed-order}.

Note that in Table~\ref{tab:fixed-order} we display the discrepancy in
$\ln \Sigma$ rather than $\Sigma$.
For effects that set in only from second order, the discrepancy in
the $\abar^2 L^2$ term is the same for both $\ln \Sigma$ and $\Sigma$.
However for effects at $\abar^3 L^3$ the discrepancies in the two
quantities differ and it is $\ln \Sigma$ that is the appropriate one to
consider. 
%
Note, further, that the thrust result has been obtained specifically
within the approximation discussed in Appendix~\ref{sec:expansion}.
Our further detailed analysis of the thrust in dipole-type showers has revealed a non-trivial interplay
between NNLL$_{\ln\Sigma}$ logarithmic terms and factors $(\as L^2)^n$, with subtleties
related to the breaking of standard exponentiation.
The nature of this interplay with the full shower evaluation of
NNLL$_{\ln\Sigma}$ terms (rather than just the recoil considerations
of Appendix~\ref{sec:expansion}) goes beyond the scope of this article.
  
\begin{table}[h]
\begin{center}
\begin{tabular}{ c c }
 Observable &  NLL$_{\ln\Sigma}$ discrepancy
  \\
\midrule
  $\sqrt{y_{3}^{\rm cam}}$  & $-0.18277\,\abar^2 L^2$\\
\cmidrule(){1-2}
   ${\rm FC}_1$      & $-0.066934\,\abar^2 L^2$\\ 
\cmidrule(){1-2}
  $B_T$              & $-0.0167335\,\abar^2 L^2$\\
\cmidrule(){1-2}
  vector $p_t$ sum   & $-0.250\,\abar^3 L^3$ \\
\cmidrule(){1-2}
  $1-T$        & $+0.016^{+0.001}_{-0.001}\,\abar^3 L^3$ \\
\cmidrule(){1-2}
\end{tabular}
\caption{The table summarises the NLL difference between the \pythia
  and \dire  shower-algorithm
  results and the analytic resummation for different observables, at
  the first non-trivial perturbative order in
  $\abar L = 2 C_F\alpha_s L/\pi$, with $L = \ln 1/v$.
  The uncertainty in the thrust case corresponds to the statistical integration
  error.
  See main text and Appendix~\ref{sec:expansion} for further details, including caveats
  concerning the thrust case.
 }
\label{tab:fixed-order}
\end{center}
\end{table}
\section{Conclusions}
\label{sec:concl}

In this article we have set out some of the formalism needed to
address the question of the multi-scale accuracy of parton showers,
specifically in the context of the {\tt Pythia} and {\tt Dire} transverse-momentum
ordered showers.
Our conclusions apply to both.

The showers essentially demonstrate all required conditions when
considering a single emission, i.e.\ the first emission is generated
in a way that reproduces the correct soft and collinear limits of the
single-emission matrix element, including single logarithmic regions
(i.e.\ large-angle soft and hard collinear splitting).
However, the pattern of multiple emission that
they generate has flaws in singular regions that are arguably
serious.
First we have found that there are double logarithmic regions,
already from two emissions, where the matrix element is incorrect at
subleading $\nc$.
This causes the subleading $\nc$ terms of the leading double logarithms
(LL$_\Sigma$) to be wrong for a number of simple and widely used
observables, such as the thrust (and $n\ge 3$ jet rates).\footnote{For
  some practitioners the surprise might be that there do exist
  observables, such as the jet broadenings and 2-jet rate, for which
  the LL$_\Sigma$ answer is correct including its subleading-$\nc$
  terms.
  We believe that at double logarithmic level, the subleading colour
  issue is relatively straightforward to fix.
  One option is to appropriately split each dipole into regions of
  $C_F$ and $C_A/2$, possibly with continuous transitions between
  them.
  The assignment would simply follow the parton identification that is
  used in angular ordered showers such as Herwig or equivalently
  outlined in Ref.~\cite{Gustafson:1992uh}.
  Note however that such an approach has a bookkeeping cost within any
  shower that at leading $N_C$ uses colour dipoles.
  In particular, any dipole, whether $q\bar q$, $qg$ or $gg$ can in
  general have an arbitrarily large number of $C_F$ and $C_A/2$
  regions, in an alternating sequence.
  This is not necessarily the only approach that one can envisage.
  Indeed other approaches have been proposed in
  Refs.~\cite{Friberg:1996xc,Giele:2011cb}.
  We prefer therefore not to advocate one or other fix for the
  subleading-colour issue without detailed studies of performance and
  computational complexity.  }

At leading-$\nc$, we have found that the effective double-emission
matrix element is wrong in a region where the two emissions have
commensurate transverse momenta and disparate angles.
This is illustrated in Fig.~\ref{fig:double-soft-ME-ratio} showing
that there are logarithmically enhanced regions where the
discrepancies are at the $100\%$ level.
This finding should perhaps not be surprising given the matrix-element
versus shower comparisons performed long ago by the Lund
group~\cite{Andersson:1991he} for transverse-momentum ordered showers
with dipole-local recoil.
The underlying characteristic in the shower algorithm that causes
this, namely the specific attribution of recoil, leads to the
NLL$_{\ln \Sigma}$ terms being wrong for a wide range of event-shape
like observables (independently of any aspects related to the CMW
scheme for the strong coupling), though the coefficient of the error
is modest.
Given the broad similarities in choices made by other
$p_\perp$-ordered dipole-type and antenna showers with local recoil, it
would not be surprising if similar conclusions apply to those as well.
The analysis methods that we have developed here provide some first
elements of a set
of tools for parton-shower authors to analyse and understand the
logarithmic properties of their algorithms.

Our observations have a number of implications.
1) NLL discrepancies (whether those observed here for two specific
transverse-momentum ordered showers, or those discussed in
Ref.~\cite{Banfi:2006gy} for angular-ordered showers) have the
potential to affect prospects
for precision physics in many of the experimental measurements that rely
significantly on parton showers.
2) The large discrepancy in the two-emission matrix elements for
the transverse-momentum ordered showers studied here, may matter
also in the field of jet substructure, where large gains
in signal to background discrimination rely on the ability to exploit
the pattern of correlations between emissions, notably with the help
of machine learning.
3) Certain methods for matching parton showers with fixed-order
calculations are made significantly more difficult if the singularity
structure of matrix elements is incorrect.
This would notably be the case for any extension of the MC@NLO method
to NNLO.
4) Efforts to improve parton showers with higher-order splitting
kernels would probably be most appropriately pursued within a
framework that is free of the issues that we encountered here.
In particular, while we discussed problems that arise for two
emissions, the underlying causes of those problems will affect matrix
elements for any number of emissions.
The inclusion of higher-order corrections to splitting functions and
of double-soft emission matrix elements might, we imagine, at best
postpone the first order in $\as$ at which the all-order logarithmic
issues first manifest themselves, but this remains a question that
deserves further study.

Overall, the approach we have introduced here provides some of the
insight needed to address the problem of how to systematically go
about creating parton shower algorithms with controlled multi-scale
accuracy.

\section*{Acknowledgements}

We are grateful to Stefan H\"oche, Paolo Nason, 
Torbj\"orn Sj\"ostrand, Peter Skands, Gregory Soyez, Bryan Webber and
Giulia Zanderighi for helpful discussions and comments on the
manuscript.
We also thank a referee for bringing
references~\cite{Andersson:1991he,Friberg:1996xc,Eden:1998ig} to our
attention and for helpful comments.
The work of P.F.M.\ has been supported by a Marie Sk\l{}odowska Curie
Individual Fellowship of the European Commission's Horizon 2020
Programme under contract number 702610 Resummation4PS.
F.D. is supported by the SNF grant P2SKP2\_165039 and by the Office of
High Energy Physics of the U.S. Department of Energy (DOE) under grant
DE-SC-0012567.
K.H.\ was supported by the European Commission through the ERC
Consolidator Grant HICCUP (No. 614577).
K.H.\ also thanks the Science and Technology Facilities
Council (STFC) for support via grant award ST/P000274/1.
M.D.\ thanks the STFC for support via grant award ST/P000800/1 and the
CERN theoretical physics department for a scientific associateship and
for hospitality during the course of this work. M.D. also acknowledges
the University of Manchester's School of Physics and Astronomy for
sabbatical leave which facilitated this work. G.P.S. and P.F.M would
like to thank the Munich Institute for Astronomy and Particle Physics
(MIAPP) for hospitality and support during the 
\emph{Automated, Resummed and Effective} programme.

\appendix
\section{Evaluation of double-soft effective matrix element}
\label{sec:app:double-soft-ME}

There are several rapidity regions that can be considered when
evaluating the double-soft effective matrix element.
Let us consider a situation where $\ln p_{\perp,1} / p_{\perp,2} \sim
\order{1}$ and where $\ln Q/p_{\perp,1} \gg 1$.
From the point of view of the identification of different rapidity
regions, we will allow ourselves inaccuracies on the rapidity of
$\order{1}$.
In particular we will consider ratios such as $\eta / \ln
(Q/p_{\perp})$, where it is immaterial whether $p_{\perp}$  is
$p_{\perp,1}$ or $p_{\perp,2}$.

\begin{figure}
  \centering
  \includegraphics[width=0.48\textwidth]{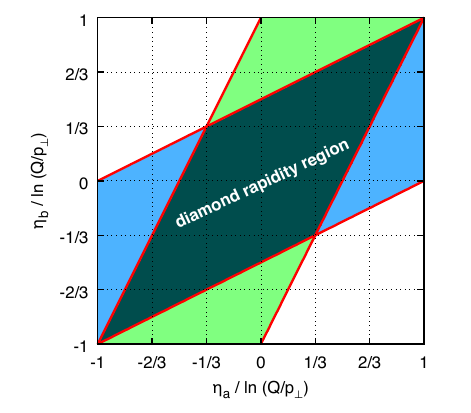}
  \caption{Representation of the different regions of rapidities for
    a pair of emissions with commensurate but small $p_\perp$ values.
    The ``diamond'' rapidity region is that used for the analysis in
    this appendix and also in Fig.~\ref{fig:double-soft-ME-ratio}.
  }
  \label{fig:two-rap-region}
\end{figure}

Figure~\ref{fig:two-rap-region} shows the two dimensions associated
with the rapidity variables, $\eta_a$, $\eta_b$ for the two emissions
$a$ and $b$, each normalised to the maximum accessible rapidity,
$\ln (Q/p_{\perp})$.
If emission $a$ comes first in the parton shower, then in the
blue-shaded region,
\begin{equation}
  \label{eq:rap-a-first}
  \frac12\left(\eta_a - \ln \frac{Q}{p_\perp} \right)
  < \eta_b
  < \frac12\left(\eta_a + \ln \frac{Q}{p_\perp} \right)
\end{equation}
emission $b$ will modify the transverse momentum of emission $a$.
Conversely, if $b$ comes first in the parton shower, then in the
green-shaded region,
\begin{equation}
  \label{eq:rap-b-first}
  \frac12\left(\eta_b - \ln \frac{Q}{p_\perp} \right)
  < \eta_a
  < \frac12\left(\eta_b + \ln \frac{Q}{p_\perp} \right)
\end{equation}
emission $a$ will modify the transverse momentum of emission $b$.
In the overlap, diamond-shaped region, it is guaranteed that the
second emission will always modify the transverse momentum of the
first one, regardless of which of $a$ and $b$ was the first one.
This is the region that we select for detailed analysis of the matrix
element.
Each of the blue and green shaded regions occupies half the plane and
the diamond overlap region occupies $1/3$ of the plane.
The region not affected at all by transverse recoil issues is also
$1/3$ of the plane.

Recall that the correct double-soft matrix element is given by
Eq.~(\ref{eq:two-emissions-correct}).
For the purpose of understanding the effective matrix element, we work
in a fixed-coupling limit.
Then consider the case where a shower generates first an emission
$g_1$ with transverse momentum ${\boldsymbol p}_{\perp,1}$
and rapidity $\eta_1$ and then an emission $g_2$ with transverse momentum ${\boldsymbol p}_{\perp,2}$ and rapidity $\eta_2$,
with the first emission's transverse momentum being modified by the
recoil to become ${\boldsymbol p}_{\perp,1} - {\boldsymbol
  p}_{\perp,2}$.
Then we evaluate the probability for either of $g_1$ and $g_2$ to
coincide with some momentum $p_a$ and other one to coincide with some
$p_b$.
This is given by:
\begin{multline}
  \label{eq:dipole-mat2}
  \frac{dP_{2,\text{shower}}(p_{a}, p_{b} \in\diamond)}{
    d\eta_a\, d\eta_b
    \;
    d^2\boldsymbol{p}_{\perp,a}
    d^2\boldsymbol{p}_{\perp,b}
  } =
  \frac1{2!} \left(\frac{\as C_A}{2\pi^2}\right)^2
  \;
  \int \frac{d^2{\boldsymbol p}_{\perp,1}}{{p_{\perp,1}}^2} 
  \int_{p_{\perp,2}<p_{\perp,1}} 
  \frac{d^2{\boldsymbol p}_{\perp,2}}{{p_{\perp,2}}^2}
  \int_{\diamond} d\eta_1 d\eta_2
  \;\times
  \\
  \times
  \left[
    \delta^2({\boldsymbol p}_{\perp,1} - {\boldsymbol p}_{\perp,2} - \boldsymbol p_{\perp,a})\,
    \delta^2({\boldsymbol p}_{\perp,2} - \boldsymbol p_{\perp,b})
    \delta (\eta_a - \eta_1) \delta (\eta_b - \eta_2)
    + (a \leftrightarrow b)
  \right]\,.
\end{multline}
Note that we do not here consider the effect of the change of rapidity
in Eqs.~(\ref{eq:recoil-impact}), because the rapidity distributions
of the gluons are uniform throughout most of the bulk of the diamond
region and a shift in rapidity leaves those uniform distributions
unchanged.
The only exceptions are at the edge of the diamond region and along
the diagonal (when the two rapidities are similar), and those regions'
phase space is suppressed by one power of the logarithm of
$v_1 \simeq v_2$.
Figure~\ref{fig:double-soft-ME-ratio} shows the ratio of
Eq.~(\ref{eq:dipole-mat2}) to the correct result,
Eq.~(\ref{eq:two-emissions-correct}), in the large $\nc$ limit, i.e.\
equating $C_F$ and $C_A/2$.

\section{Fixed-order difference with respect to NLL resummation}
\label{sec:expansion}
In this appendix we report the necessary formulae to carry out the
third-order study of the difference $\delta\Sigma(L)$ between the dipole showers
considered in this work and the NLL analytic result, the results of
which are summarised in Table~\ref{tab:fixed-order}.

There are a number of simplifications that one can make in organising
the calculation.
A first simplification comes from the fact that the distribution of
the first emission (prior to any of the subsequent emissions) is
correctly described by the dipole showers considered here, and
therefore the difference $\delta\Sigma(L)$ starts at
${\cal O}(\abar^2)$. Secondly, we are only interested in
configurations in which the real emissions are simultaneously soft and
collinear, and widely separated in rapidity from each other, which
contribute to $\Sigma(L)$ starting at NLL. The correct emission
probability in these configurations amounts to
\begin{equation}
  \label{eq:n-emissions-correct}
  dP_{n} = \frac{C_F^n}{n!} \prod_{i=1}^n \left (
  \frac{2\as(p_{\perp,i}^2) }{\pi} \frac{dp_{\perp,i}}{p_{\perp,i}}
  d\eta_i \frac{d\phi_i}{2\pi}\right)\,,
\end{equation}
where $p_{\perp,i}$ and $\eta_i$ are defined with respect to the $q$
and $\bar q$ directions and
\begin{equation}
\label{eq:rapidity_bound_sc}
|\eta_i| \lesssim \ln\frac{Q}{p_{\perp,i}}\,.
\end{equation}
Eq.~(\ref{eq:n-emissions-correct}) is valid as long as
the emissions are very separated in rapidity, even if they have
commensurate $p_{\perp,i}$.

A third observation is that, in the soft and collinear limit, the
ordering variables of the dipole showers studied in this work coincide
with the transverse momentum in Eq.~\eqref{eq:n-emissions-correct},
hence we write $p_{\perp,i}=v_i$ in the following. Moreover, given
that we are focusing on a fixed-order comparison, we can safely ignore
running coupling effects and set
\begin{equation}
 C_F \frac{2\as(p_{\perp}^2) }{\pi} \to C_F \frac{2\as(Q^2) }{\pi} = \abar\,.
\end{equation}
We stress that all considerations made in this section are strictly
valid in the large-$N_c$ limit, where we equate $C_A=2C_F$.

We then start by considering a double-emission configuration, ordered
in the transverse momenta $v_i$. In a shower picture, in addition to
the real-emission probabilities, one needs to include the contribution
of the no-emission probability between the hard scale $Q=1$ and the
scale $v_1$ at which the first emission occurs, given by the following
Sudakov form factor
\begin{equation}
\label{eq:radiator}
e^{-R(v_1)} \equiv \exp\left\{ - \abar\int_{v_1}^{1} \frac{d p_{\perp}}{p_{\perp}}
  \int_{-\ln\frac{1}{p_\perp}}^{\ln\frac{1}{p_\perp}} d\eta \int_{0}^{2\pi}\frac{d\phi}{2\pi}\right\}.
\end{equation}
Equivalent suppression factors account for the no-emission probability
between $v_1$ and $v_2$ and between $v_2$ and the shower cutoff,
$v_i > Q_0$. One can expand the Sudakov factors out at fixed order and
take the limit $Q_0\to 0$. This allows one to write all virtual
corrections explicitly and obtain a fixed-order expansion of the
shower equation.

Since the single-emission event is correctly described by the dipole
showers, all single and double-virtual corrections at ${\cal
  O}(\abar^2)$ cancel (to NLL accuracy) in the
difference $\delta\Sigma(L)$, which at this order is fully determined
by the following double-real contribution
\begin{multline}
  \label{eq:as2-expansion}
  \delta \Sigma^{\rm (2\,emissions)}(L) = 
  \abar^2
  \int_0^1 \frac{dv_1}{v_1}
  \int_{\ln v_1}^{\ln 1/v_1} d\eta_1
  \int_0^{v_1} \frac{dv_2}{v_2}
  \int_{\ln v_2}^{\ln 1/v_2} d\eta_2
  \int_0^{2\pi}
  \frac{d\phi_{1}}{2\pi} \int_0^{2\pi}\frac{d\phi_{2}}{2\pi} 
  \times
  \\
  \times
  \left[
    \Theta\big(e^{-L} - V(p_1^\text{shower},p_2)\big)
    - \Theta\left(e^{-L} - V(p_1^\text{correct},p_2)\right)
  \right],
\end{multline}
where we traded the $1/2!$ multiplicity factor for the ordering
$v_1\geq v_2$. 

At the next non-trivial order we need to add configurations with three
real emissions, for which one can repeat the above derivation
obtaining
\begin{align}
  \label{eq:as3-expansion}
  \delta \Sigma^{\rm (3\,emissions)}&(L) = 
  \abar^3
  \int_0^1 \frac{dv_1}{v_1}
  \int_0^{v_1} \frac{dv_2}{v_2}\int_0^{v_2} \frac{dv_3}{v_3}
  \int_{\ln v_1}^{\ln 1/v_1} d\eta_1\int_{\ln v_2}^{\ln 1/v_2} d\eta_2 \int_{\ln v_3}^{\ln 1/v_3} d\eta_3\times\notag\\
& \times\int_0^{2\pi}\frac{d\phi_{1}}{2\pi}\int_0^{2\pi}\frac{d\phi_{2}}{2\pi}  \int_0^{2\pi}\frac{d\phi_{3}}{2\pi}\, \times\notag\\
&  \times \Big[\Theta(e^{-L} -
  V(p_1^\text{shower},p_2^\text{shower},p_3)) - \Theta(e^{-L} -
  V(p_1^\text{correct},p_2^\text{correct},p_3)) \notag\\
&-  \Theta(e^{-L} -
  V(p_1^\text{shower},p_2)) + \Theta(e^{-L} -
  V(p_1^\text{correct},p_2))\notag\\
&- \Theta(e^{-L} - V(p_1^\text{shower},p_3)) + \Theta(e^{-L} -
  V(p_1^\text{correct},p_3)) \notag\\
& -\Theta\big(e^{-L} - V(p_2^\text{shower},p_3)\big)
    + \Theta\left(e^{-L} - V(p_2^\text{correct},p_3)\right)
  \Big].
\end{align}
The labels ``$\text{correct}$'' and ``$\text{shower}$'' in the momenta
of Eqs.~\eqref{eq:as2-expansion} and~\eqref{eq:as3-expansion} indicate
that the emissions' momenta are modified according to the recoil
prescription as either in the correct NLL result or in the dipole
shower, respectively. While in the correct solution the recoil for the
considered phase-space configurations is always absorbed by the
emitting Born leg (either $q$ or $\bar q$), as discussed in
Section~\ref{sec:double-emission-case-single} this is not the case for
the dipole showers analysed here. Instead, given $n-1$ emissions and
the Born legs $q$ (along the positive $z$ direction) and $\bar q$
(along the negative $z$ direction), the recoil for the $n$-th emission
$p_n$ can be assigned according to the following cases (as in the main
text, we denote with a tilde all quantities prior to the emission of
$p_n$):
\begin{itemize}
\item $p_n$ is emitted off the $[\bar q g_i]$ dipole (with $p_i$ being
  the momentum of the gluon $g_i$ colour-connected to the anti-quark
  $\bar q$):
  \begin{align}
  \label{eq:B7}
  {\rm if}&\,\,  \eta_n > \frac{1}{2}\left(\ln \widetilde{p}_{\perp,
      i}+\widetilde{\eta}_i\right):~~ \boldsymbol{p}_{\perp,
      i}= \widetilde{\boldsymbol{p}}_{\perp,
      i} - \boldsymbol{p}_{\perp,
      n};~~\eta_i= \widetilde{\eta}_i - \ln\frac{|\widetilde{\boldsymbol{p}}_{\perp,
      i} - \boldsymbol{p}_{\perp,
      n}|}{|\widetilde{\boldsymbol{p}}_{\perp,
      i} |}\,,\\
    {\rm if}&\,\,  \eta_n < \frac{1}{2}\left(\ln \widetilde{p}_{\perp,
      i}+\widetilde{\eta}_i\right):~~ \boldsymbol{p}_{\perp,
      \bar q}= \widetilde{\boldsymbol{p}}_{\perp,
      \bar q} - \boldsymbol{p}_{\perp,
      n}.
  \end{align}
\item  $p_n$ is emitted off the $[g_i q]$ dipole (with $p_i$ being
  the momentum of the gluon $g_i$ colour-connected to the quark $q$):
  \begin{align}
  {\rm if}&\,\,  \eta_n < \frac{1}{2}\left(-\ln \widetilde{p}_{\perp,
      i}+\widetilde{\eta}_i\right):~~ \boldsymbol{p}_{\perp,
      i}= \widetilde{\boldsymbol{p}}_{\perp,
      i} - \boldsymbol{p}_{\perp,
      n};~~\eta_i= \widetilde{\eta}_i + \ln\frac{|\widetilde{\boldsymbol{p}}_{\perp,
      i} - \boldsymbol{p}_{\perp,
      n}|}{|\widetilde{\boldsymbol{p}}_{\perp,
      i} |}\,,\\
    {\rm if}&\,\,  \eta_n > \frac{1}{2}\left(-\ln \widetilde{p}_{\perp,
      i}+\widetilde{\eta}_i\right):~~ \boldsymbol{p}_{\perp,
      q}= \widetilde{\boldsymbol{p}}_{\perp,
      q} - \boldsymbol{p}_{\perp,
      n}.
  \end{align}
 \item  Finally, if $p_n$ is emitted off a $[g_i g_j]$ dipole (with
   $\widetilde{\eta}_i < \widetilde{\eta}_j$):
  \begin{align}
  {\rm if}&\,\,  \eta_n < \frac{1}{2}\left(\ln \frac{\widetilde{p}_{\perp,
      j}}{\widetilde{p}_{\perp,
      i}}+\widetilde{\eta}_i + \widetilde{\eta}_j\right):~~ \boldsymbol{p}_{\perp,
      i}= \widetilde{\boldsymbol{p}}_{\perp,
      i} - \boldsymbol{p}_{\perp,
      n};~~\eta_i= \widetilde{\eta}_i + \ln\frac{|\widetilde{\boldsymbol{p}}_{\perp,
      i} - \boldsymbol{p}_{\perp,
      n}|}{|\widetilde{\boldsymbol{p}}_{\perp,
      i} |}\,,\\
 {\rm if}&\,\,  \eta_n > \frac{1}{2}\left(\ln \frac{\widetilde{p}_{\perp,
      j}}{\widetilde{p}_{\perp,
      i}}+\widetilde{\eta}_i + \widetilde{\eta}_j\right):~~ \boldsymbol{p}_{\perp,
      j}= \widetilde{\boldsymbol{p}}_{\perp,
      j} - \boldsymbol{p}_{\perp,
      n};~~\eta_j= \widetilde{\eta}_j - \ln\frac{|\widetilde{\boldsymbol{p}}_{\perp,
      j} - \boldsymbol{p}_{\perp,
      n}|}{|\widetilde{\boldsymbol{p}}_{\perp,
      j} |}.
  \label{eq:B12}
\end{align}
\end{itemize}
In the virtual corrections corresponding to the $\Theta$ functions
with only two emissions in Eq.~\eqref{eq:as3-expansion}, the recoil
procedure must be applied only to the momenta that are used in the
corresponding observable, hence ignoring the third (virtual) momentum,
which is insensitive to recoil.

\begin{figure}
  \centering
  \includegraphics[width=0.7\textwidth]{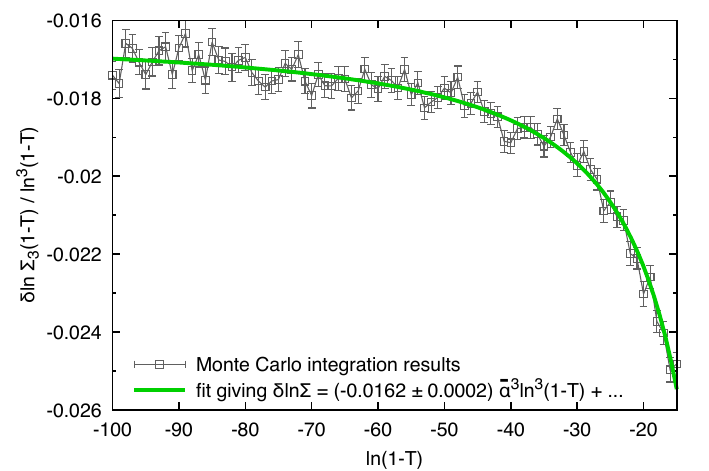}
  \caption{Evaluation of $\delta \ln \Sigma_3$ (defined here as the
    coefficient of $\abar^3$ rather than $\as^3$ as elsewhere in the
    text) for the thrust, $T$, as a function of $\ln(1-T)$, within the
    approximation of Eqs.~(\ref{eq:B7})--(\ref{eq:B12}), and divided
    by $\ln^3(1-T)$ to help visualise the coefficient of
    $\abar^3 \ln^3(1-T)$.
    The fit function assumes a form for $\delta \ln \Sigma_3$ that is
    a third-order polynomial in powers of $\ln(1-T)$.
    The corresponding result for the coefficient of $\abar^3 L^3$ in
    Table~\ref{tab:fixed-order} displays a larger, somewhat
    conservative error.
    This is to account for the dependence of the fit
    result on the precise choice of fit range, as well as potential
    correlations between points, which are not taken into account in the
    calculation of the error from the fit.}
  \label{fig:thrust-fit}
\end{figure}

The above formulae can be used to numerically check our analytical
results, or directly evaluate results in cases where we do not yet
have an analytical answer.
This approach has been used for all results in
Table~\ref{tab:fixed-order}. 
%
%
Note that the table shows $\delta \ln \Sigma$ rather than the
$\delta\Sigma$ as evaluated in Eqs.~(\ref{eq:as2-expansion})
and (\ref{eq:as3-expansion}).
Writing $\Sigma = 1 + \sum_{n=1}^\infty \as^n \Sigma_n$, up to
third order we  have
$\ln \Sigma = \as \Sigma_1 + \as^2 \left(\Sigma_2-\frac12 \Sigma_1^2\right) +
\as^3 \left(\Sigma_3 - \Sigma_1 \Sigma_2 + \frac13\Sigma_1^3 \right) +
\order{\as^4}$.
The evaluation of $\delta \ln \Sigma_n$ in particular requires
evaluation of differences $\delta \Sigma_m$ for $m \le n$.
In practice, considering solely the effects of recoils as parametrised
in Eqs.~(\ref{eq:B7})--(\ref{eq:B12}), differences start only at
second order, i.e.\ $\Sigma_2$ onwards.
Keeping this is mind, one obtains
$\delta \ln \Sigma = \as^2 \delta\Sigma_2 + \as^3(\delta\Sigma_3 -
\!\Sigma_1 \delta\Sigma_2) + \order{\as^4}$.
In practice it can be numerically advantageous to evaluate the
expansion of $\delta \ln \Sigma$ directly.
This is shown for the thrust in Fig.~\ref{fig:thrust-fit}.

For the vector $p_t$ sum we have carried out an analytical evaluation
and it gives
\begin{equation}
  \delta \ln \Sigma_3 =
  -\frac{1}{2} \int_0^1 \frac{d\zeta}{\zeta}\int_0^{2\pi}
  \frac{d\phi}{2\pi} \ln^2(1 + \zeta^2 + 2\zeta \cos\phi)
  + 7 g = -\frac{\zeta(3)}{2} + 7g \simeq -0.250\,,
  \label{eq:vecpt-result}
\end{equation}
where $g$ is given by
\begin{multline}
  \label{eq:g-constant}
  g =
  \int_0^1 \frac{d\zeta_1}{\zeta_1} \int_0^{\zeta_1}
  \frac{d\zeta_2}{\zeta_2}
  \int_0^{2\pi} \frac{d\phi_1}{2\pi} \int_0^{2\pi}
  \frac{d\phi_2}{2\pi}
  \times \\ \times
  \frac12 \ln \left( 1 + \zeta_1^2 + \zeta_2^2 + 2\zeta_1 \cos\phi_1 +
    2\zeta_2 \cos\phi_2 + 2\zeta_1 \zeta_2 \cos(\phi_1 - \phi_2)\right)
  \simeq 0.050\,.
\end{multline}
We suspect, but have not proven, that $g=\zeta(3)/24$.
Eq.~(\ref{eq:vecpt-result}) has been verified by direct numerical
evaluation of Eqs.~(\ref{eq:as2-expansion}) and
(\ref{eq:as3-expansion}).

%
\bibliographystyle{JHEP}
\bibliography{MC}

\end{document}